\documentclass[lettersize,journal]{IEEEtran}
\usepackage{amsmath,amsfonts}
\usepackage{algorithmic}
\usepackage{algorithm}
\usepackage{array}
\usepackage[caption=false,font=normalsize,labelfont=sf,textfont=sf]{subfig}
\usepackage{textcomp}
\usepackage{stfloats}
\usepackage{url}
\usepackage{verbatim}
\usepackage{graphicx}
\usepackage{cite}
\usepackage{xcolor}
\usepackage[normalem]{ulem}

\usepackage{multirow}
\usepackage{booktabs}
\begin{document}

\title{Decentralized Proactive Model Offloading and Resource Allocation for Split and Federated Learning}

\author{
    Binbin Huang,~\IEEEmembership{Member,~IEEE}, 
    Hailian Zhao, 
    Lingbin Wang,
    Wenzhuo Qian,
    Yuyu Yin,~\IEEEmembership{Member,~IEEE},
    Shuiguang Deng ~\IEEEmembership{Senior Member,~IEEE}
\thanks{Binbin Huang, Lingbin Wang and Yuyu Yin are with the School of Computer, Hangzhou Dianzi University, Hangzhou, China. \protect  E-mail: huangbinbin@hdu.edu.cn, 232320030@hdu.edu.cn, yinyuyu@hdu.edu.cn.}
\thanks{Hailiang Zhao, Wenzhuo Qian, Shuiguang Deng are with the School of Computer Science and Technology, Zhejiang University. E-mail: hliangzhao@zju.edu.cn, qwz@zju.edu.cn, dengsg@zju.edu.cn.}
\thanks{Hailiang Zhao is the corresponding author.}
\thanks{Copyright (c) 20xx IEEE. Personal use of this material is permitted. However, permission to use this material for any other purposes must be obtained from the IEEE by sending a request to pubs-permissions@ieee.org.}
}

\markboth{Journal of \LaTeX\ Class Files,~Vol.~14, No.~8, August~2021}%
{Shell \MakeLowercase{\textit{et al.}}: A Sample Article Using IEEEtran.cls for IEEE Journals}


\maketitle

\begin{abstract}
In the resource-constrained IoT-edge computing environment, Split Federated (SplitFed) learning is implemented to enhance training efficiency. This method involves each terminal device dividing its full DNN model at a designated layer into a device-side model and a server-side model, then offloading the latter to the edge server. However, existing research overlooks four critical issues as follows: (1) the heterogeneity of end devices' resource capacities and the sizes of their local data samples impact training efficiency; (2) the influence of the edge server's computation and network resource allocation on training efficiency; (3) the data leakage risk associated with the offloaded server-side sub-model; (4) the privacy drawbacks of current centralized algorithms. Consequently, proactively identifying the optimal cut layer and server resource requirements for each end device to minimize training latency while adhering to data leakage risk rate constraint remains a challenging issue. To address these problems, this paper first formulates the latency and data leakage risk of training DNN models using Split Federated learning. Next, we frame the Split Federated learning problem as a mixed-integer nonlinear programming challenge. To tackle this, we propose a decentralized Proactive Model Offloading and Resource Allocation (DP-MORA) scheme, empowering each end device to determine its cut layer and resource requirements based on its local multidimensional training configuration, without knowledge of other devices' configurations. Extensive experiments on two real-world datasets demonstrate that the DP-MORA scheme effectively reduces DNN model training latency, enhances training efficiency, and complies with data leakage risk constraints compared to several baseline algorithms across various experimental settings.

\end{abstract}

\begin{IEEEkeywords}
IoT-Edge Computing, Model Offloading, Resource Allocation, Data Leakage Risk, Decentralized Algorithm.
\end{IEEEkeywords}

\section{Introduction}
\IEEEPARstart {V}{arious} Deep Neural Networks (DNNs) have facilitated tremendous progress across a wide smart Internet of Things (IoT) applications, such as intelligent transportation \cite{liu2022energy,otoum2022feasibility}, smart healthcare \cite{nguyen2022federated,ali2022federated} and smart home \cite{rasti2022graph}. In these smart IoT applications, large volumes of data generated by end devices is often private and sensitive. To utilize sensitive data to train DNN models in a safe manner, federated Learning (FL) as a privacy preserving machine learning technique, is introduced\cite{li2021survey,konevcny2016federated}. FL collaborates multiple end devices to train a DNN model in a distributed manner while keeping data locally. In the FL framework, each end device loads a full DNN model and parallelly trains its local DNN model based on local data samples, and then aggregates its local model to form a global model in an edge server. Thus, FL can obtain a DNN model without exposing sensitive data, thereby greatly satisfying data privacy requirement. However, due to the limited resource capacities of end devices, federated learning can suffer from two major problems: (1) end device with limited memory capacity could not afford to run the DNN model whose total training memory footprint exceeds the memory capacity of a single end device; (2) The mismatch between weak computation capacity of the end device and prohibitive computation workload of DNN model training leads IoT-oriented FL to time-consuming and ineffective.
To address these two problems, split federated (SplitFed) learning \cite{thapa2022splitfed,turina2021federated} is introduced to split each end device's full DNN model at a cut layer into a device-side model and a server-side model and offload the server-side model to the resource-adequate edge server and achieve device-edge synergy training, consequently alleviating the aforementioned two problems.

Research problem about utilizing SplitFed learning to enable IoT-oriented FL to run on resource-constrained end devices and further improve training efficiency of IoT-oriented FL has attracted much attention in academia \cite{thapa2022splitfed,wu2023split,jiang2022fedsyl,wu2022fedadapt,samikwa2022ares,turina2020combining}. However, existing studies mainly exist a few major problems:

(1) Most studies have failed to consider the heterogeneity of end devices' resource capacities and their local data samples size on the training efficiency. If allocating the same sub-model to each end device, the weaker end device with more local data samples need to spend more time in training models, inevitably prolonging the total training latency and reducing the training efficiency of FL. For example, the authors in \cite{thapa2022splitfed} and \cite{turina2020combining} split the DNN model on each end device at the same cut layer and train the DNN model in parallel or sequentially.

(2) Most studies have failed to consider the impact of the edge server's computation and network resource allocation on the training efficiency. In device-edge synergy training, multiple resource-constrained end devices connected to an edge server for distributed DNN model training share network resources and computing resources of the edge server. An efficient resource allocation can further reduce the training latency and improve the training efficiency. Otherwise, the total training latency can be prolonged.

(3) Most studies have failed to consider the data privacy leakage problem. Existing studies offload server-side sub-model to the edge server from resource-constrained end devices, thereby accelerating the DNN model training. However, it is possible to recover raw sample data from the knowledge of the gradient parameter of offloaded server-side sub-model \cite{zhu2019deep, geiping2020inverting}. Hence, it can incur data privacy leakage when offloading server-side sub-model to the server for device-edge synergy training. Therefore, it is necessary to design an appropriate model offloading strategy to trade off the computation efficiency of FL and its data leakage risk. 

(4) Most studies adopt centralized algorithms to identify the optimal cut layer for end devices with different resource capacities. The formulation and solution of the centralized optimization problem itself is a detriment on privacy. That is because centralized algorithms are generally built on the complete knowledge regarding all end devices' multiple dimensional training configurations,
including the computation capacity, the mini-batch size, the size of the local dataset and the number of epochs, etc. Hence, it may not be possible in many real-world smart applications exploiting SplitFed learning. It is more realistic for each end device to decide the cut layer and the required resources according to its local multiple-dimensional training configurations.

This paper aims to address the above problems, with a particular focus on reducing the overall training time while satisfying the data leakage risk rate constraint. We first present a system architecture. We then formulate the latency and data leakage risk of DNN model training adopting split federated learning. The DNN model training process consists of multiple rounds, each of which consists of \emph{Starting Phase}, \emph{Intermediate Phase} and \emph{End Phase} three stages. We characterize the latency for these three stages and the total training latency in one training round, respectively. We next adopt a data-driven methodology to fit the forward/backward propagation workloads, the smashed data/the smashed data’s gradient size, and the device-side model’s data size as functions of the cut layer, respectively. Based on these, we formulate the split federated learning problem as a mixed integer non-linear programming. To solve this problem, we design a decentralized proactive model offloading and resource allocation (DP-MORA) scheme which enables each end device to decide its own cut layer and resource requirement according to its own private information without knowing other end devices' private information. Finally, extensive experimental results on real-world datasets demonstrate that compared with several baseline algorithms, the newly proposed DP-MORA scheme can reduce the total training latency while satisfying the data leakage risk rate constraint. The main contributions of this paper are summarized as follows:

\begin{itemize}
    \item Based on massive prior experiments, we measure the data leakage risk rates with respect to different cut layers. The data leakage risk rate is measured by the cosine similarity between the local data samples and the data samples recovered from the server-side model. The larger cosine similarity, the higher data leakage risk rate.

    \item We formulate joint model offloading and resource allocation problem for split federated learning to be a mixed integer non-linear programming problem, aiming at minimizing the training latency while satisfying data leakage risk rate constraint.

    \item We propose a novel DP-MORA scheme. In the scheme, each end device can decide its own model cut layer, radio spectrum allocation and computation resource allocation according to its local multiple-dimensional training configurations, without knowing the multiple-dimensional training configurations of other end devices. The related private information of other end devices can be fully preserved.
\end{itemize}

The rest of this paper is organized as follows. Section 2 reviews the related works. Section 3 presents the system model. Section 4 formulates joint model offloading and resource allocation problem as a mixed integers non-linear programming problem. Section 5 presents a novel DP-MORA scheme in detail. Section 6 conducts \sout{the} extensive experiments and evaluates the performance of DP-MORA scheme. Finally, section 7 concludes the paper and points out future work.

\section{Related work}

In a resource-constrained IoT-edge computing environment, split federated learning enables each resource-constrained end device to split its local full DNN model into the device-side model and server-side model at the cut layer and offload the server-side model to the edge server, thereby achieving device-edge synergy training and improving training efficiency. How to identify the optimal cut layer to accelerate model training in a resource-constrained IoT-edge computing environment has attracted much attention in academia. There are a lot of studies about exploiting split federated learning to accelerate DNN model training in a distributed manner. These studies mainly can be classified into two types: efficiency-oriented split federated learning and efficiency and privacy-oriented split federated learning.

\subsection{Efficiency-oriented split federated learning}
For efficiency-oriented device-edge synergy training, the authors \cite{abedi2023fedsl} proposed a novel federated split learning framework to efficiently train models on distributed sequential data. Analogously, the authors in \cite{thapa2022splitfed} designed two schemes, called SplitFedv1 and SplitFedv2, to amalgamate split learning and federated learning two approaches, and sequentially train the client/server-side models. These two schemes accelerate local training in resource-constrained end devices by offloading partial layers of the DNN model to the edge server. However, the work manually determines the same cut layer for each end device's local DNN model and has failed to consider the heterogeneity of end devices' resource capacities. To address the shortcomings of the identical sub-models on heterogeneous end devices, the authors \cite{wu2022fedadapt} in view of the computational heterogeneity and changing network bandwidth, adopted reinforcement learning to adaptively identify the optimal cut layer for each end device's local DNN model. Its main goal is to reduce the training latency. Analogously, the authors in \cite{bakhtiarnia2023dynamic} designed an approach to dynamically determine the optimal cut layer according to the state of the communication channel, and thereby improving the efficiency of model inference. The authors \cite{samikwa2022ares} proposed an adaptive resource-aware split-learning scheme in IoT systems to trade off the training latency and energy consumption. Moreover, the authors in \cite{han2021accelerating} proposed a local-loss-based split learning to optimally split the model and train the client/server-side models in parallel, aiming at reducing the training latency. Although the aforementioned works adaptively identify the cut layers for multiple heterogeneous end devices to improve the training efficiency, they have failed to take the data leakage risk into account. Therefore, these solutions cannot be directly applied to data privacy requirement IoT-edge computing environment.

\subsection{Efficiency and privacy-oriented split federated learning}
For efficiency and privacy-oriented device-edge synergy training, the authors in \cite{jiang2022fedsyl} pre-train a regression model to identify the optimal cut layer and evaluate the privacy leakage risk rate by server-side model parameter quantities. To trade off the training efficiency and the data leakage risk, a federated synergy learning paradigm is proposed. However, the work only uses the device-side model size to measure the privacy leakage, which lacks effective demonstration in real data reconstruction attacks. To cope with this problem, the authors in \cite{deng2023hsfl} proposed a metric called inverse efficiency to measure privacy leakage. Based on this, they formulated model decomposition with privacy constraints in hybrid split learning and federated learning as a constraint optimization problem and transformed it into a contextual bandit problem. To address this problem, an efficient contextual bandit learning-based scheme is developed to identify the optimal cut layers for end devices, aiming to optimize the training latency and data privacy protection. The authors in \cite{turina2021federated} proposed a novel learning architecture that combines split learning and federated learning two approaches to improve the training efficiency and data privacy. Analogously, the authors in \cite{zhang2023privacy} proposed a new hybrid Federated Split Learning architecture to trade off privacy protection and the training latency. Moreover, the authors in \cite{pham2023binarizing} adopted a binarized split learning to reduce the end device's computation workload and memory usage. To further preserve privacy, they integrated differential privacy into the binaries split learning model and trained it with additional local leak loss. Its main goal is to lightweight model and preserve privacy. Although the aforementioned works have jointly considered the cut layer selection and privacy leakage, there is still no work to consider the impact of the edge server's resource allocation on the training efficiency. Moreover, the aforementioned works adopted the centralized algorithms to identify the optimal cut layer of each individual end device. The formulation of the centralized optimization problem itself is a detriment to privacy. 
 
Inspired by the above motivations, we propose a metric called data leakage risk rate to measure the data leakage incurred by different selections of cut layer. Based on it, we jointly consider model offloading and resource allocation problem for split and federated learning in IoT-edge computing environment. To cope with this problem, we design a decentralized proactive model offloading and resource allocation scheme, aiming to minimize the training latency while satisfying the data leakage risk rate constraints.

\section{System model}
\subsection{System architecture}
\begin{figure}[h]
    \centering
    \includegraphics[scale=0.16]{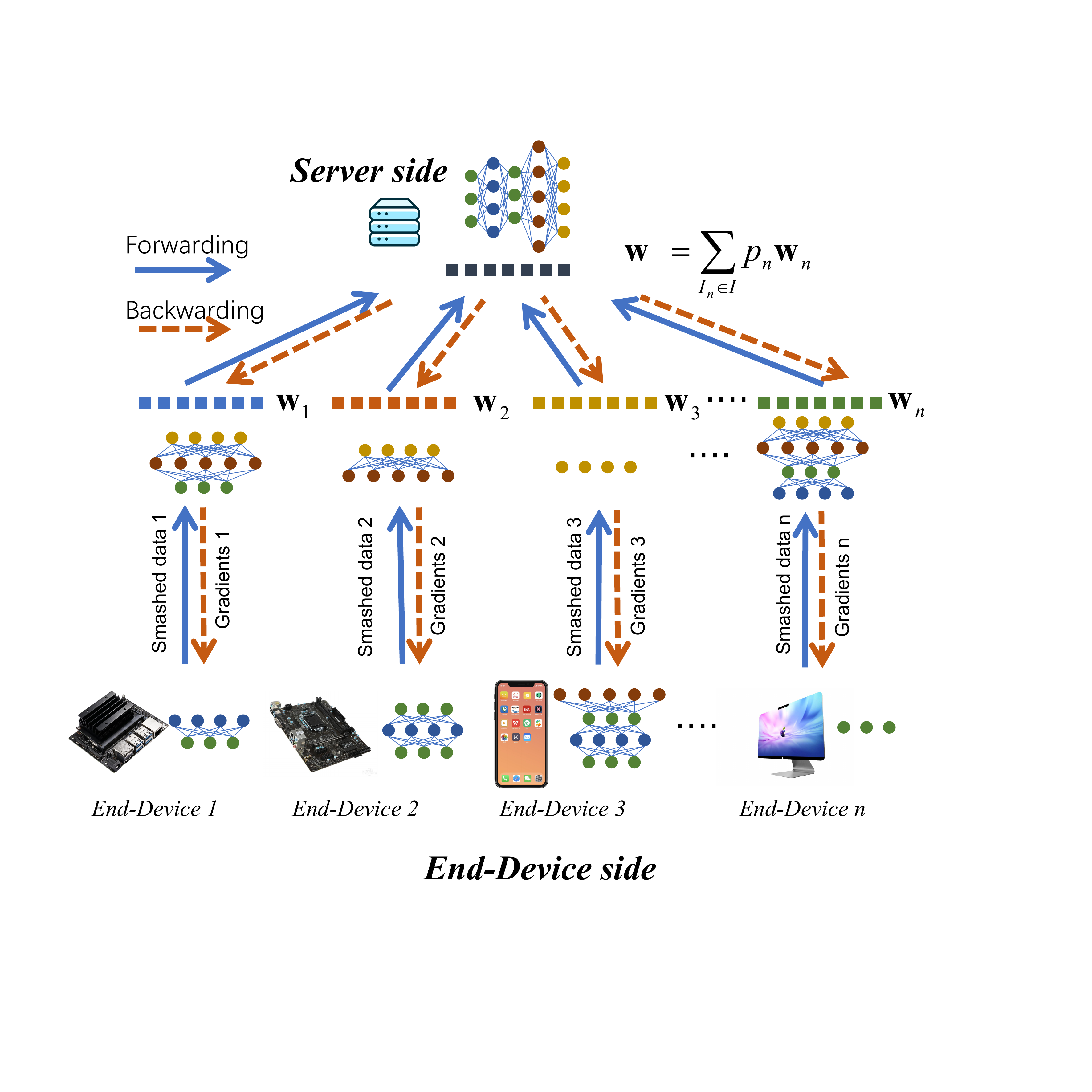}
    \vspace{-2.4cm}
    \caption{System architecture for split federated learning}
    \label{fig:fig1_architecture}
\end{figure}

As illustrated in Fig. \ref{fig:fig1_architecture}, we consider a device-edge synergy paradigm comprising of an edge server and $N$ heterogeneous end devices. These $N$ end devices are resource constraint and the computing capacity of each end device $I_{n}$ is $f_{d}^{n}$. The edge server is resource adequate and its computing capacity is $f_{s}$. By exploiting split federated learning, the edge server can cooperate with the $N$ resource-constrained end devices to train a global DNN model in a distributed manner without revealing their sample data to the edge server. The DNN model is composed of $L$ consecutive layers, each denoted by $l$, with $l\in\{1,\cdots,L\}$. The layer is denoted as the fundamental component of a DNN model. The set of $N$ end devices can be denoted by $\mathcal{I}=\{I_1,\cdots,I_n,\cdots,I_N\}$. Each end device $I_n$ owns its local dataset $\mathcal D_n =\{\textbf{z}_n^i,y_n^i\}_{i=1}^{D_n}$, the size of which is denoted as $D_n = |\mathcal D_n|$. Here, $\textbf{z}_n^i \in \mathbb R^{1\times Q}$ and $y_i\in \mathbb R^{1\times 1}$ represent an input data sample and its corresponding label, respectively, where $Q$ denotes the dimension of the input data sample. The aggregated dataset over all end devices is represented by $\mathcal D=\bigcup_{n=1}^N\mathcal D_n$.  A local DNN model on each end device $I_n$ is split into a device-side sub-model and a server-side sub-model at a cut layer $l_n=\alpha_nL, \alpha_n\in [0,1], l_n\in \{0,\cdots,L\}$. The intermediate output associated with the cut layer is called \emph{smashed data}. The device-side sub-model deployed on end device $I_n$ is denoted by $\mathbf {w}_d^n$. The server-side sub-model deployed on edge server is denoted by $\mathbf {w}_s^n$. The DNN model on end device $I_n$ is denoted by $\mathbf{w}_n=\{\mathbf{w}_d^n;\mathbf {w_s^n}\}$. At each training round, the edge server cooperates these $N$ end devices to train their models in parallel for multiple epochs, and then aggregate their latest device-side models to the edge server to obtain the updated global DNN model. It is worth noting that the cut layer can not be the input layer for device data privacy preservation consideration. A special case is that cut layer $l_n=L$ means an empty server-side model. In other words, the device-edge synergy training degrades to the federated learning scheme with $N$ end devices. The whole training process can be divided into $\mathcal{T}=\{1,\cdots,t,\cdots,T\}$ rounds. Key symbols and their descriptions used in this paper are summarized in Table I.

\begin{table}[!t]
\caption{Symbols and Their Definitions}
\label{regressionModels}
\centering
\begin{tabular}{m{1cm}|m{7cm}}
\toprule
Symbol  & Definition \\
\midrule
$f_{d}^n$ & The $nth$ end device’s computing capacity\\
$f_{s}$ & The edge server's computing capacity\\
$l_n$ & Denotes the $nth$ model’s cut layer \\
$D_{n}$ & The $nth$ end device's local dataset \\
$\Upsilon $ & The local train epochs at round t \\
$w_d^n$ & The nth device-side sub-model \\
$w_s^n$ &  The nth server-side sub-model \\
$r_{d,n}^{DL}$ & The downlink transmission rate from the edge server to end device $I_n$ \\
$\tau_{s,n}^{m, DL}$ & The distribution latency of device-side sub-model   $w_d^n$ for end device $I_n$ \\
$\tau_{s,n}^{f,e}$ &The computing latency of device-side model processing forward propagation on a mini-batch of data samples \\
$\tau_{d,n}^{s,UL}$ & The smash data transmission latency from end device to edge server\\
$\tau_{s,n}^{f,e}$ &The computing latency of server-side model processing a forward propagation on a single data sample\\
$\tau_{s,n}^{b,e}$ &The computing latency of server-side model processing a backward propagation on a single data sample\\
$\tau_{s,n}^{g,DL}$ &The smash data’s gradient transmission latency for a mini-batch data samples\\
$\tau_{d,n}^{b,e}$ &The computing latency of device-side model processing backward propagation on a mini-batch of data samples\\
$\tau_{n}^{t}$ & The overall training latency of end device $I_n$ in one training round t  \\
$b_{d}^{n}$ & The total batches of end device $I_n$ for an epoch\\
$\tau_{n,s}^{m,UL}$ & The transmission latency for each end device transmits the device-side model to edge server \\
$\mathbf{S}_d^n(t,1)$ & The smashed data gets from the end device executes its device-side model with the raw data samples \\
$P_d^n(l_n)$ & The data leakage risk  of cut layer $l_n$ for each end device $I_n$  \\
$\theta_d^n$ & The ratio of server's computing capacity allocated to the $nth$ end device  \\
$\mu_{d,n}^{DL/UL}$ & The ratio of communication capacity allocated to the $nth$ end device\\
$\alpha_n$ & The factor for nth end device dividing the model into device-side model and server-side model \\
\hline
\end{tabular}
\end{table}

\subsection{The latency of DNN model training adopting split federated learning}
As a new device-edge synergy paradigm, the split federated learning is adopted to train a global DNN model with $T$ rounds. Each round consists of \emph{Starting Phase}, \emph{Intermediate Phase} and \emph{End Phase} three stages. We analyze the latency for these three stages in detail. Based on these, we can further characterize the overall training latency as follows.

\emph{(1) Starting Phase}. The starting phase mainly includes device-side model distribution. At the beginning of training round $t$, the latest device-side sub-model $\mathbf{w}_d^n(t)$ is distributed to the end device $I_n$ by the edge server. The device-side sub-model's size (in bits) for each end device $I_n$ is $\psi_{d,n}^m(l_n)$, depending on cut layer $l_n$. The downlink transmission rate $r_{d,n}^{DL}$ from the edge server to end device $I_n$ can be calculated by

\begin{equation}              r_{d,n}^{DL}=\mu_{d,n}^{DL}W^{DL}log_2(1+P_{s}|h_d^n|^2/W^{DL}N_0), \forall I_n \in \mathcal{I}  
\end{equation}
where $\mu_{d,n}^{DL}$ is the time fraction allocated to the end device $I_n$ on the downlink channel. $W^{DL}$ is the radio spectrum bandwidth for the edge server’s uplink channel. $P_{s}$ is the edge server's transmission power. $h_d^n$ is the channel gain between the edge server and end device $I_n$. $N_0$ is the thermal noise spectrum density. Hence, the distribution latency $\tau_{d,n}^{m,DL}$ of device-side sub-model $\mathbf{w}_d^n$ for end device $I_n$ can be calculated by
\begin{equation}\label{startPhase}
    \tau_{s,n}^{m,DL}=\psi_{d,n}^m(l_n)/r_{d,n}^{DL}
\end{equation}

\emph{(2) Intermediate Phase}. After receiving its latest device-side model, each end device $I_n$ trains its DNN model $\mathbf{w}_n$ for $\Upsilon$ epochs in training round $t$, indexed by $ \upsilon \in \{1,\cdots,\upsilon,\cdots,\Upsilon\}$. At the first epoch in training round $t$, we have $\mathbf{w}_d^n(t,1)\leftarrow \mathbf{w}_d^n(t), \forall I_n \in \mathcal{I}$. For each end device $I_n$, its model training process at each epoch $\upsilon$ in training round $t$ consists of device-side model execution, smashed data transmission, server-side model execution, server-side model update, smashed data's gradient transmission and device-side model update six parts. The latency of each part can be analyzed as follows.

\emph{• Device-side model execution latency}. It represents the time taken by the device-side model $\mathbf{w}_d^n$ of end device $I_n$ performing forward propagation on a mini-batch data samples. At an epoch $\upsilon$ in training round $t$, each end device $I_n$ randomly draws a mini-batch of data samples $\mathcal{B}_d^n$ from its local dataset $\mathcal{D}_d^n$ and feed them into device-side model to perform forward propagation. Here, the mini-batch size for end device $I_n$ can be denoted by $B_d^n=|\mathcal{B}_d^n|$. Hence, the total batches $b_d^n$ of end device $I_n$ for an epoch is calculated by $b_d^n=D_d^n/B_d^n$. Let $\phi_{s,n}^{f,e}(l_n)$ denote the computation workload (in FLOPs) for the device-side model of end device $I_n$ performing forward propagation on a single data sample. Hence, the latency $\tau_{d,n}^{f,e}$ for the device-side model processing a mini-batch of data samples can be calculated by

\begin{equation}        
 \tau_{d,n}^{f,e}=(B_d^n\phi_{d,n}^{f,e}(l_n))/f_d^n, \forall I_n \in \mathcal{I} 
\end{equation}

\emph{• Smashed data transmission latency}. It represents the time taken by end device $I_n$ transmitting the \emph{smashed data} for a mini-batch of data samples to the edge server. After the device-side model performing forward propagation on a mini-batch of data samples, the \emph{smashed data} of these data samples need to be transmitted to the edge server using the allocated bandwidth resource. Let $\psi_{d,n}^{s,UL}(l_n)$ be the smashed data size of one data sample at the cut layer $l_n$. Hence, the smashed data size in bits for a mini-batch $B_d^n$ of data samples can be denoted by $B_d^n\psi_{d,n}^{s,UL}(l_n)$. The number of subcarriers allocated to end device $I_n$ can be denoted by $\mu_d^n$. The uplink transmission rate $r_{d,n}^{UL}$ from end device $I_n$ to edge server can be calculated by

\begin{equation}        
 r_{d,n}^{UL}=\mu_{d,n}^{UL}W^{UL}log_2(1+P_d^n|h_d^n|^2/W^{UL}N_0), \forall I_n \in \mathcal{I} 
\end{equation}
where $W^{UL}$ is the radio spectrum bandwidth for the edge server's downlink channel. $P_d^n$ denotes the transmission power for end device $I_n$. $h_d^n$ denotes the channel gain between end device $I_n$ and edge server. Hence, the smash data transmission latency is given by 

\begin{equation}        
 \tau_{d,n}^{s,UL}=(B_d^n\psi_{d,n}^{s,UL}(l_n))/r_{d,n}^{UL}(t,\upsilon), \forall I_n \in \mathcal{I} 
\end{equation}

\emph{• Server-side model execution latency}. It represents the time taken by the server-side model $\mathbf{w}_s^n$ of end device $I_n$ performing forward propagation on a mini-batch data samples. Let $\phi_{s,n}^{f,e}(l_n)$ denote the computation workload for the server-side model $\mathbf{w}_s^n$ of end device $I_n$ performing forward propagation on a single data sample. Since the smash data for a mini-batch of data samples are fed for training the server-side model, the overall computation workload is $B_d^n\phi_{s,n}^{f,e}(l_n)$. Hence, the server-side model execution latency can be calculated by

\begin{equation}        
 \tau_{s,n}^{f,e}=(B_d^n\phi_{s,n}^{f,e}(l_n))/(\theta_d^n f_s), \forall I_n \in \mathcal{I} 
\end{equation}
where $f_s$ is the computation capacity of the edge server. $\theta_d^n$ denotes the edge server's computation capacity fraction allocated to end device $I_n$.

\emph{• Server-side model update latency}. It represents the time taken by the server-side model $\mathbf{w}_s^n$ of end device $I_n$ performing backward propagation on a mini-batch data samples. Let $\phi_{s,n}^{b,e}(l_n)$ denote the computation workload for the server-side model $\mathbf{w}_s^n$ of end device $I_n$ performing backward propagation on a single data sample. Hence, the overall computation workload for the server-side model $\mathbf{w}_s^n$ of end device $I_n$ performing backward propagation on a mini-batch of data samples is $B_d^n\phi_{s,n}^{b,e}(l_n)$. The server-side model update latency can be calculated by

\begin{equation}        
 \tau_{s,n}^{b,e}=(B_d^n\phi_{s,n}^{b,e}(l_n))/(\theta_d^n f_s), \forall I_n \in \mathcal{I} 
\end{equation}

\emph{• Smashed data's gradient transmission latency}. It represents the time taken by the edge server transmitting the smashed data's gradients for a mini-batch of data samples to end device $I_n$. After server-side model of end device $I_n$ is updated, its smashed data's gradients are sent back to end device $I_n$ using the allocated radio spectrum. Let $\psi_{s,n}^{g,DL}(l_n)$ denote the data size of smashed data's gradient for a single data sample. Hence, the smash data's gradient transmission latency for a mini-batch data samples can be calculated by

\begin{equation}        
 \tau_{s,n}^{g,DL}=(B_d^n\psi_{s,n}^{g,DL}(l_n))/r_{d,n}^{DL}, \forall I_n \in \mathcal{I} 
\end{equation}

\emph{• Device-side model update latency}. It refers to the time taken by the device-side model $\mathbf{w}_d^n$ of end device $I_n$ performing backpropagation on a mini-batch data samples. Let $\phi_{d,n}^{b,e}(l_n)$ represent the computation workload for the device-side model $\mathbf{w}_d^n$ of end device $I_n$ performing backward propagation on a single data sample. Hence, the device-side model update latency can be calculated by 

\begin{equation}
    \tau_{d,n}^{b,e}=(B_d^n\phi_{d,n}^{b,e}(l_n))/(f_d^n), \forall I_n \in \mathcal{I}
\end{equation}
Based on the above analysis, the overall latency of end device $I_n$ at epoch $\upsilon$ in training round $t$ can be calculated by

\begin{equation}\label{intermediatePhase}
   \tau_{n}^{m,e}=b_d^n(\tau_{d,n}^{f,e}+\tau_{d,n}^{s,UL}+\tau_{s,n}^{f,e}+\tau_{s,n}^{b,e}+\tau_{s,n}^{g,DL}+\tau_{d,n}^{b,e}), \forall I_n \in \mathcal{I}
\end{equation}

\emph{(3) End Phase}. The end phase is to aggregate the latest device-side models to the edge server to obtain the updated global DNN model. Therefore, the latency for the end phase is mainly consisting of the device-side model transmission latency and the model parameters aggregation latency on the edge server. After each end device $I_n$ training its DNN model $\mathbf{w}_n$ consisting of device-side model $\mathbf{w}_d^n$ and server-side model $\mathbf{w}_s^n$ for $\Upsilon$ epochs, the latest device-side model is transmitted from each end device $I_n$ to the edge server, and the corresponding transmission latency $\tau_{n,s}^{m,UL}$ for each end device $I_n$ can be calculated by 

\begin{equation}\label{endPhase}
    \tau_{n,s}^{m,UL}=\psi_{d,n}^m(l_n)/r_{d,n}^{UL}, \forall I_n \in \mathcal{I}
\end{equation}
where $\psi_{d,n}^m(l_n)$ is the device-side sub-model size (in bits) for each end device $I_n$. $r_{d,n}^{UL}$ is the uplink transmission rate from the end device to the edge serve. The model parameters are aggregated on the edge server by the FedAvg algorithm. Due to its low computational complexity, the model aggregation latency is very small and negligible.

\emph{(4) Overall training latency}. With the results of three phases in (\ref{startPhase}), (\ref{intermediatePhase}) and (\ref{endPhase}), the overall training latency of end device $I_n$ in one training round $t$ can be calculated by

\begin{equation}
\tau_n^t(l_n,\mu_{d,n}^{DL},\mu_{d,n}^{UL},\theta_d^n )=\tau_{s,n}^{m,DL}+\Upsilon\tau_{n}^{m,e}+\tau_{n,s}^{m,UL}, \forall I_n \in \mathcal{I}
\end{equation}
which depends on the cut layer $l_n$ of DNN model on end device $I_n$, the time fraction $\mu_{d,n}^{DL}$ allocated to end device $I_n$ on the downlink channel, the time fraction $\mu_{d,n}^{UL}$ allocated to end device $I_n$ on the uplink channel, and the edge server's computation capacity's fraction $\theta_d^n$ allocated to end device $I_n$. Specifically, the cut layer affects not only the computation workload distribution and the transmission data amount between the end device and the edge server but also the data leakage risk rate. Different cut layers have different amounts of transmission data. A shallow cut layer means light computational workload on end devices, but high data leakage risk. A deep cut layer means heavy computation workloads on end devices, but low data leakage risk. We further quantify the data leakage risk in next subsection.

\subsection{The data leakage risk of DNN model training adopting split federated learning}
Although offloading model from resource-constrained end devices to the edge server can reduce the training latency and improve training efficiency, the raw data can be partially reconstructed from the gradient information \cite{geiping2020inverting,zhu2019deep}, thereby incurring the data leakage risk. The recovery of data samples from gradient information is described in detail as follows:

\emph{(1) Server-side model's gradient of the original data samples}. In the first epoch of training round $t$, the edge server owns the device-side model $\mathbf{w}_d^n(t,1)$ of each end device $I_n$. Let $\mathbf{Z}_d^n(t,1)\in \mathbb{R}^{B_d^n\times Q}$ denote the aggregated input of a mini-batch of data samples in end device $I_n$. Each end device $I_n$ executes its device-side model with the raw data samples, and obtains smashed data $\mathbf{S}_d^n(t, 1) \in \mathbb{R}^{B_d^n\times P}$, i.e.,

\begin{equation}
    \mathbf{S}_d^n(t,1)=f(\mathbf{Z}_d^n(t,1);\mathbf{w}_d^n(t,1)), \forall I_n \in \mathcal{I} 
\end{equation}
where $f(\mathbf{Z}_d^n;\mathbf{w}_d^n)$ represents the mapping function between a mini-batch of data samples $\mathbf{Z}_d^n$ and their smash data $\mathbf{S}_d^n$ given device-side sub-model parameter $\mathbf{w}_d^n$. $P$ denotes the dimension of smashed data for one data sample. Device-side model $\mathbf{w}_d^n$ of each end device $I_n$ performs forward propagation on the aggregated input $\mathbf{Z}_d^n(t,1)$ of a mini-batch of data samples and transmits its smashed data $\mathbf{S}_d^n(t,1)$ to the edge server. The edge server receives the smashed data $\mathbf{S}_d^n(t,1)$ of each end device $I_n$ and feeds it into the server-side model $\mathbf{w}_s^n(t,1)$. As such, the predicted result from the server-side model is given by 

\begin{equation}
    \mathbf{\hat y}_d^n(t,1)=f(\mathbf{S}_d^n(t,1);\mathbf{w}_s^n(t,1))\in\mathbb R^{B_d^n\times1}, \forall I_n \in \mathcal{I} 
\end{equation}
According to the loss function between the predicted results $\mathbf{\hat y}_d^n$ and the corresponding ground-truth labels $\mathbf{y}_d^n$, the gradient of the server-side model can be calculated and denoted by $\nabla L(\mathbf{w}_s^n(t,1))$.

\emph{(2) Server-side model's gradient of the recovered data samples}. Let $\mathbf{Z}_{d,n}^{'}(t,1)\in \mathbb{R}^{B_d^n\times Q}$ denote the mini-batch of recovered data samples in end device $I_n$. The edge server executes the device-side model of end device $I_n$ with the mini-batch of recovered data samples, and obtains corresponding smashed data $\mathbf{S}_{d,n}^{'}(t, 1) \in \mathbb{R}^{B_d^n\times P}$

\begin{equation}
    \mathbf{S}_{d,n}^{'}(t, 1)=f(\mathbf{Z}_{d,n}^{'}(t,1);\mathbf{w}_d^n(t,1)), \forall I_n \in \mathcal{I}
\end{equation}
 
The smashed data $\mathbf{S}_{d,n}^{'}(t, 1)$ is fed into the server-side model $\mathbf{w}_s^n(t,1)$ of end device $I_n$, and it outputs the predicted results $\hat{\mathbf{y}}_{d,n}^{'}(t,1)$,

\begin{equation}
    \mathbf{\hat y}_{d,n}^{'}(t,1)=f(\mathbf{S}_{d,n}^{'}(t,1);\mathbf{w}_s^n(t,1))\in\mathbb R^{B\times1}, \forall I_n \in \mathcal{I} 
\end{equation}
According to the loss function between the predicted results $\mathbf{\hat y}_{d,n}^{'}$ and the corresponding ground-truth labels $\mathbf{y}_d^n$, the server-side model's gradient of the recovered data samples can be calculated and denoted by $\nabla L^{'}(\mathbf{w}_s^n(t,1))$.

\emph{(3) Recovering data samples from their gradients}. In the first epoch of each training round $t$, the edge server owns the device-side model $\mathbf{w}_d^n(t,1)$ and the server-side model $\mathbf{w}_s^n(t,1)$. Thus, the edge server trys to recover data samples by optimizing an euclidean matching term between $\nabla L(\mathbf{w}_s^n(t,1))$ and $\nabla L^{'}(\mathbf{w}_s^n(t,1))$ \cite{geiping2020inverting,zhu2019deep}. The optimization objective function

\begin{equation}
    arg\min\limits_{\mathbf{Z}_{d,n}^{'}(t,1)} 1-\frac{<\nabla L(\mathbf{w}_s^n(t,1)),\nabla L^{'}(\mathbf{w}_s^n(t,1))>}{||\nabla L(\mathbf{w}_s^n(t,1))||||\nabla L^{'}(\mathbf{w}_s^n(t,1))||}
\end{equation}
is minimized to recover the original data samples $\mathbf{Z}_d^n(t,1)\in \mathbb{R}^{B_d^n\times Q}$ from a gradient $\nabla L(\mathbf{w}_s^n(t,1))$. 

According to the above analysis, we define the data leakage risk $P_d^n(l_n)$ of cut layer $l_n$ for each end device $I_n$ as the cosine similarity of the original data samples $\mathbf{Z}_d^n(t,1)\in \mathbb{R}^{B_d^n\times Q}$ and the recovered data samples $\mathbf{Z}_{d,n}^{'}(t,1)\in \mathbb{R}^{B_d^n\times Q}$,

\begin{equation}
    P_d^n(l_n)=\frac{<\mathbf{Z}_d^n(t,1),\mathbf{Z}_{d,n}^{'}(t,1)>}{||\mathbf{Z}_d^n(t,1)||||\mathbf{Z}_{d,n}^{'}(t,1)||}
\end{equation}

\subsection{Regression-based modeling methodology}
As shown in Subsection 3.2, in the overall training latency model, some functions including the forward propagation workload function with respect to the cut layer, the backward propagation workload function with respect to the cut layer, the smashed data size function with respect to the cut layer, the smashed data's gradient's size with the respect to the cut layer, the device-side model's data size with respect to the cut layer, cannot be expressed in an analytic form. This is because these functions vary with different DNN models. For example, the specific coefficients in the above functions are different for ResNet 18 and ResNet 34 two models.

To address the above challenges, referring to the existing works \cite{wang2022leaf+,jiang2022fedsyl}, we adopt a data-driven methodology. We first empirically measure the forward/backward propagation workloads, smashed data/smashed data's gradient sizes, and device-side model's data size with respect to different cut layers for ResNet 18 and ResNet 34 two models. Then we employ a variety of classic regression models to profile the above relationships. We find that it is optimal to adopt Quadratic Polynomial Regression (QPR) regression models to profile the functions between the forward/backward propagation workloads, the device-side model size and the cut layer. Also, it is optimal to adopt Reciprocal Regression (RR) to profile the function between the smashed data/smashed data's gradient size and the cut layer. The developed regression models for ResNet 18 and ResNet 34 two models are shown in Table \ref{regressionModels}.

The root means square error (RMSE) is applied for calculating the average model-prediction error in the units of the variable of interest \cite{willmott2005advantages}.

\begin{table}[!t]
\caption{The Proposed Regression-based Models}
\label{regressionModels}
\centering
\begin{tabular}{m{1.2cm}|c|c|c}
\toprule
DNN Models  & Functions    & Proposed models   & RMSE \\
\midrule
\multirow{5}{*}{ResNet 18}  &  $\psi^m(l_n)$   & $0.9746x^2-5.58x+6.528$  & 3.235\\
 & $\phi^{f,e}(l_n)$   & $-0.01597x^2+0.7705x-0.4282$   & 0.115\\
 & $\phi^{b,e}(l_n)$   & $0.01597x^2-0.7705x+5.8946$   & 0.115\\
 & $\psi^{s,UL}(l_n)$   & $3.2028/x-0.3443$   & 0.275\\
 & $\psi^{s,DL}(l_n)$   & $3.2028/x-0.3443$   & 0.275\\
\midrule
  \multirow{5}{*}{ResNet 34}  &  $\psi^m(l_n)$   & $0.4795x^2-3.517x+5.001$  & 8.242\\
 & $\phi^{f,e}(l_n)$   & $-0.00274x^2+0.7044x-0.3718$   & 0.312\\
 & $\phi^{b,e}(l_n)$   & $0.00274x^2-0.7044x+11.3978$   & 0.312\\
 & $\psi^{s,UL}(l_n)$   & $2.891/x-0.0987$   & 0.164\\
 & $\psi^{s,DL}(l_n)$   & $2.891/x-0.0987$   & 0.164\\
\bottomrule
\end{tabular}
\end{table}

\section{Problem formulation}
In a resource-constrained heterogeneous IoT-edge computing environment, considering the heterogeneity of end devices' resource capacities and their local data samples' size, it is necessary to split the local DNN model of each end device at different cut layers, and then offload different server-side models to the edge server to achieve device-edge synergy learning. In device-edge synergy learning, due to the limited network and computation resources of edge server, it is also necessary to efficiently allocate the sharing resource to further reduce the DNN training latency. Here, we formulate the joint model offloading and resource allocation problem to be the problem $P_1$. Its main goal is to reduce the total training latency for $N$ end devices in a round $t$ while satisfying data leakage risk rate constraints. We define our optimization objective and constrained conditions as follows: 

\begin{equation}
\begin{aligned}
    \label{eqP1}\mathcal{P}_1: \mathop{min} \limits_{l_n,\mu_{d,n}^{DL},\mu_{d,n}^{UL},\theta_d^n}& Q=\sum_{n=1}^{N}\tau_n^t(l_n,\mu_{d,n}^{DL},\mu_{d,n}^{UL},\theta_d^n) \\
    \mathrm{s.t. } & C_1: P_d^n(\alpha_n L)\le P^{risk}, \forall I_n \in \mathcal{I}  \\
    &C_2: \sum_{n=1}^N\mu_{d,n}^{DL}\le 1 \\
    &C_3: \sum_{n=1}^N\mu_{d,n}^{UL}\le 1 \\
    &C_4: \sum_{n=1}^N \theta_d^n\le 1 \\
    &C_5: l_n=\alpha_n L\in\left\{1,2,...,L\right\}, \forall I_n \in \mathcal{I}  \\
    &C_6: \mu_{d,n}^{DL}\in \left(0,1\right), \mu_{d,n}^{UL}\in \left(0,1\right), \\
    &  \theta_d^n\in \left(0,1\right), \forall I_n \in \mathcal{I}
\end{aligned}
\end{equation}
where $\tau_n^t(l_n,\mu_{d,n}^{DL},\mu_{d,n}^{UL},\theta_d^n )$ is the overall training latency of end device $I_n$ in a round $t$. Constraint condition $C_1$ in Eq. \ref{eqP1} is the data leakage risk rate constraint of each end device $I_n$, in which $P^{risk}$ is the risk rate constraint. Constraint condition $C_2$ is the time fraction allocation vector to $N$ end devices on the uplink channel. Constraint condition $C_3$ is the time fraction allocation vector to $N$ end devices on the downlink channel. Constraint condition $C_4$ denotes that the sum of computing resources allocated to $N$ end devices cannot exceed the computing capacity of the edge server. Constraint condition $C_5$ guarantees the cut layer of each end device $I_n$ is an integer. 



The formulation of $\mathcal{P}_1$ is a maxed-integer non-linear programming problem (MINLP) which is difficult to solve \cite{belotti2013mixed}. To solve the formulation of $\mathcal{P}_1$ in a centralized algorithm itself is a detriment on privacy. That is because the centralized algorithms are generally built on the complete knowledge regarding all end devices' multiple dimensional training configurations, including the computation capacity, the mini-batch size, the size of the local dataset and the number of epochs, etc. Hence, it may not be suitable for many real-world smart applications exploiting SplitFed Learning. It is urgent to design a novel DP-MORA scheme which enables each end device to decide its own cut layer and resource requirement according to its own multiple-dimensional training configuration without knowing other end devices’ multiple-dimensional training configurations.

\section{Proposed DP-MORA optimization algorithm}
The block coordinate descent (BCD) method is to divide the variables of the optimization problem into multiple blocks, and only optimize the variables of one block in each iteration, while keeping the other variables fixed. Gradually optimizing a block of variables can result in much less computation than directly optimizing all variables. Therefore, BCD method is suitable to solve complex optimization problems with different properties about different variables (i.e., the convex problem with some variables or non-convex problem with another variable). Moreover, BCD shows good convergence in practice. In convex optimization problems, BCD is usually able to guarantee global convergence, while in non-convex problems, BCD is able to converge to a local optimal solution. In many practical problems, BCD can guarantee to converge to a local optimal solution by gradually optimizing the variable blocks. The local optimization of BCD on each subproblem makes the objective function value gradually decrease. Although the global optimum is not guaranteed, a better solution is usually obtained. Our formulated problem $P_1$ is a complex optimization problem with multiple variables. Subproblem with respect to different variables have different properties. For example, subproblem with variable $\mu_{d,n}^{DL}$ is convex problem, while the subproblem with respect to the variable $l_n$ is non-convex problem. The BCD algorithm has achieved excellent performance in solving this kind of problem. To achieve this, we propose a DP-MORA scheme based on the block coordinate descent (BCD) method \cite{grippo2000convergence}, which enables each end device to decide its own cut layer and resource requirement according to its own multiple-dimensional training configurations. We first relax the formulation of $\mathcal{P}_1$ into the formulation of $\mathcal{P}_2$, then demonstrate the designing of the DP-MORA scheme.


$\forall I_n\in \mathcal{I}$, we relax the discrete variable ${\alpha}_n$ into continuous variable $\hat{\alpha}_n$. The relaxed formulation of $\mathcal{P}_1$ is denoted as follows:
\begin{equation}
\begin{aligned}
    \mathcal{P}_2: \mathop{minimize} \limits_{\hat{\alpha}_n,\mu_{d,n}^{DL},\mu_{d,n}^{UL},\theta_d^n}& Q=\sum_{n=1}^{N}\tau_n^t(l_n,\mu_{d,n}^{DL},\mu_{d,n}^{UL},\theta_d^n)\\
    \mathrm{s.t. } & \hat{C}_1: P_d^n(\hat{\alpha}_nL)\le P^{risk}, \forall I_n \in \mathcal{I}  \\ 
    &C_2, C_3, C_4, C_6\\
    &\hat{C}_5: 1\leq\hat{\alpha}_nL\leq L, \forall I_n \in \mathcal{I}
\end{aligned}
\end{equation}

According to the BCD method, we propose the joint model offloading and resource allocation scheme which solves problem $\mathcal{P}_2$ by sequentially fixing three of four variables and updating the remaining one to solve problem $\mathcal{P}_2$. We iterate the process until the value of each variable converges.

Let $\nabla y(x)$ denote the partial derivative of function $y$ corresponding to variable $x$. Let $\text{Proj}_\mathcal{X} (x)=\text{arg} \min_{\kappa\in \mathcal{X}}||x-\kappa||^2$ denote the Euclidean projection of $x$ onto $\mathcal{X}$. The procedure of our proposed solution is presented in detail in Algorithm 1, and can be summarized as:

(1) Given $\mu_{d,n}^{DL}$, $\mu_{d,n}^{UL}$ and $\theta_d^n$, the decision variable of the optimization problem $\mathcal{P}_2$ is $\hat{\alpha}_n$. Each end device $I_n$ can derive its new $\hat{a}_n$ according to
\begin{equation}\label{hat_a_upate}
    \hat{a}_n^{j+1}=\text{Proj}_{\mathcal{X}_{\hat{a}_n}} (\hat{a}_n^j-\eta_{\hat{a}_n} \nabla Q(\hat{a}_n^j)), \forall I_n \in \mathcal{I}
\end{equation}
where $\eta_{\hat{a}_n} > 0$ is a constant step size and ${\mathcal{X}_{\hat{a}_n}}$ is the bounded domain constrained by $\hat{C}_5$ of Problem $\mathcal{P}_2$. Based on the BCD method, we repeat Eq. \ref{hat_a_upate} until the derived $\hat{a}_n$ is converged and then update $\hat{a}_n$. ($\hat{a}_n$ can converge to a local optimal solution).

(2) Given the latest value $\hat{a}_n$ updated in step (1), $\mu_{d,n}^{UL}$ and $\theta_d^n$, the decision variable of the optimization problem $\mathcal{P}_2$ is $\mu_{d,n}^{DL}$. The optimization problem $\mathcal{P}_2$ is simplified to
\begin{equation}
\begin{aligned}
    \mathcal{P}_3: \mathop{minimize} \limits_{\{\mu_{d,n}^{DL}, \forall I_n \in \mathcal{I}\}}& Q=\sum_{n=1}^{N}\tau_n^t(\mu_{d,n}^{DL}) \\
    \mathrm{s.t. } &C_2: \sum_{n=1}^N\mu_{d,n}^{DL}\le 1 ,\\
    &C_6: \mu_{d,n}^{DL}\in \left(0,1\right), \mu_{d,n}^{UL}\in \left(0,1\right), \\
    &  \theta_d^n\in \left(0,1\right), \forall I_n \in \mathcal{I}
\end{aligned}
\end{equation}
where constraints $\hat{C}_1, C_3, C_4, \hat{C}_5$ are irrelevant to this problem. Due to the time fraction allocation constraint $C_2$ on the downlink channel shared by $N$ end devices, these $N$ end devices are coupled with each other, which leads to the method for solving $\hat{\alpha}_n$ not applied for solving $\mu_{d,n}^{DL}$. To address the optimization problem $\mathcal{P}_2$, a decentralized proactive downlink bandwidth resource allocation scheme is designed, the procedure of which is presented in detail in Algorithm 2. In order to eliminate the coupling relationship of $N$ end devices incurred by constraint $C_2$ and make $N$ end devices independently make decision its $\mu_{d,n}^{DL}$, the constraint $C_2$ can be further rewritten as
\begin{equation}
\begin{aligned}
    \mathcal{P}_3: \mathop{minimize} \limits_{\{\mu_{d,n}^{DL}, \forall I_n \in \mathcal{I}\}}&Q=\sum_{n=1}^{N}\tau_n^t(\mu_{d,n}^{DL}) \\
    \mathrm{s.t. } &C_2: \sum_{n=1}^N\mu_{d,n}^{DL}\le \sum_{n=1}^N\frac{1}{N} ,\\
    &C_6: \mu_{d,n}^{DL}\in \left(0,1\right), \mu_{d,n}^{UL}\in \left(0,1\right), \\
    &  \theta_d^n\in \left(0,1\right), \forall I_n \in \mathcal{I}
\end{aligned}
\end{equation}

By introducing the Lagrangian multipliers $\lambda$ for $C_2$, the Lagrangian function of the optimization problem $\mathcal{P}_2$ is

\begin{equation}   q_n(\mu_{d,n}^{DL},\lambda_n)=\tau_n^t(\mu_{d,n}^{DL})-\lambda_n\mu_{d,n}^{DL}+\lambda_n(1/N)
\end{equation}
Thus, the Lagrangian duality of $C_2$ with multiplier $\lambda$ is defined as
\begin{equation}
\begin{aligned}
    &\mathop{max} \limits_{\lambda}q(\lambda) \\
    =&\mathop{max} \limits_{\lambda} \sum_{\{\forall I_n \in \mathcal{I}\}}q_n(\lambda_n) \\
    =&\mathop{max} \limits_{\lambda} \inf_{\mu_{d,n}^{DL}\in (0,1)} (\sum_{\{\forall I_n \in \mathcal{I}\}} (\tau_n^t(\mu_{d,n}^{DL})-\lambda_n\mu_{d,n}^{DL}+\lambda_n(1/N)))
\end{aligned}
\end{equation}

If using the above duality directly, the subproblem needs to be solved centrally to calculate the gradients at the edge server due to the global multiplier $\lambda$. To avoid this shortage, a constrained optimization problem with Laplacian matrix $L$ and local multiplier vector $\mathbf{\Lambda}=\emph{col}(\lambda_1,\ldots, \lambda_n,\ldots,\lambda_N)$ is formulated as

\begin{equation}\label{max_Q}
\begin{aligned}
    &\mathop{max} \limits_{\Lambda}Q(\Lambda)=\mathop{max} \limits_{\Lambda}\sum_{\{\forall I_n \in I\}}q_n(\lambda_n) \\
    &\mathrm{s.t. } C_1: L\Lambda=0
\end{aligned}
\end{equation}
where the constraint $C_1$ is to guarantee $\lambda_1=\ldots=\lambda_n=\ldots=\lambda_N$, which represents that the consensus is reached. Based on the above reformulations, the optimization problem $P_3$ can be decoupled and each end device can decide its own time fraction allocations on the downlink channel of edge server. 

The augmented Lagrangian duality of problem (\ref{max_Q}) with Lagrangian multipliers $\mathbf{Z}=\emph{col}(z_1,\ldots, z_n,\ldots,z_N)$ is given by

\begin{equation}\label{augmentedLagrangian}
\begin{aligned}
    &\mathop{min} \limits_{Z}\mathop{max} \limits_{\Lambda}Q(\Lambda,\mathbf{Z}) \\
    =&\mathop{min} \limits_{Z}\mathop{max} \limits_{\Lambda} (\sum_{\{\forall I_n \in I\}}q_n(\lambda_n)-\mathbf{Z}^TL\Lambda-\frac{1}{2}\Lambda^TL\Lambda)
\end{aligned}
\end{equation}

To solve the problem (\ref{augmentedLagrangian}), the gradient flow is applied in \cite{yi2016initialization}, and the gradients $\nabla\mu_{d,n}^{DL}$, $\nabla\lambda_n$ and $\nabla z_n$ for $\mu_{d,n}^{DL}$, $\lambda_n$ and $z_n$ can be calculated by 

\begin{equation}
\begin{aligned}
   \nabla\mu_{d,n}^{DL}(j+1)=\mathop{Proj}\limits_{\mathcal{X}_{\mu_{d,n}^{DL}}}& (\mu_{d,n}^{DL}(j)-\nabla \tau_n^t(\mu_{d,n}^{DL})+\lambda_n(j))-\\
   \mu_{d,n}^{DL}(j)
\end{aligned}
\end{equation}

\begin{equation}
\begin{aligned}
    &\nabla\lambda_n(j+1)=-\sum_{I_m\in \mathcal{A}_n}(\lambda_n(j)-\lambda_m(j))- \\
    &\sum_{I_m\in \mathcal{A}_n}(z_n(j)-z_m(j))+(1/N-\mu_{d,n}^{DL}(j))
\end{aligned}
\end{equation}

\begin{equation}
    \nabla z_n(j+1)=\sum_{I_m\in \mathcal{A}_n}(\lambda_n(j)-\lambda_m(j))
\end{equation}
where $\mathcal{A}_n$ denotes the set of other end devices connected to end device $I_n$. Hence, to reach the global solution, each end device $I_n$ needs to know the information on $\lambda_m$, $z_m$, $I_m\in\mathcal{A}_n$. In our scenario, each end device cannot directly communicate with each other. However, since they are all connected to the edge server, they can utilize the edge server to relay the information on $\lambda_m$, $z_m$, $I_m\in\mathcal{A}_n$ to end device $I_n$. Based on the gradients $\nabla\mu_{d,n}^{DL}$, $\nabla\lambda_n$ and $\nabla z_n$, $\mu_{d,n}^{DL}$, $\lambda_n$ and $z_n$ can be updated by

\begin{equation}
    \mu_{d,n}^{DL}(j+1)=\mu_{d,n}^{DL}(j)+\eta\nabla\mu_{d,n}^{DL}(j+1)
\end{equation}

\begin{equation}
    \lambda_n(j+1)=\lambda_n(j)+\eta\nabla\lambda_n(j+1)
\end{equation}

\begin{equation}
    z_n(j+1)=z_n(j)+\eta\nabla z_n(j+1)
\end{equation}
where $\eta$ represents the integration step. 

(3) Given the latest value $\hat{a}_n$ updated in step (1), the latest value $\mu_{d,n}^{DL}$ updated in step (2) and $\theta_d^n$, the decision variable of the optimization problem $\mathcal{P}_2$ is $\mu_{d,n}^{UL}$. Due to the time fraction allocation constraint $C_3$ on the uplink channel shared by $N$ end devices, these end devices are also coupled with each other. In addition, the problem formulation about the decision variable $\mu_{d,n}^{UL}$ is the same to that about the decision variable $\mu_{d,n}^{DL}$ in step (2). Hence, we adopt the same decentralized and privacy-preserving uplink bandwidth resource allocation scheme to solve the decision variable $\mu_{d,n}^{DL}$.

(4) Given the latest value $\hat{a}_n$ updated in step (1), the latest value $\mu_{d,n}^{DL}$ updated in step (2) and the latest value $\mu_{d,n}^{DL}$ updated in step (3), the decision variable of the optimization problem $\mathcal{P}_2$ is $\theta_d^n$. Similarly, due to the constraint $C_4$ on the edge server's computing capacity shared by $N$ end devices, these end devices are also coupled with each other. Moreover, the problem formulation about the decision variable $\theta_d^n$ is the same to that about the decision variable $\mu_{d,n}^{DL}$ in step (2). Hence, we adopt the same decentralized and privacy-preserving computation resource allocation scheme to solve the decision variable $\theta_d^n$.

\begin{algorithm}[ht]
    \caption{Joint model offloading and resource allocation strategy based BCD algorithm}
    \label{The MORA algorithm}
    \begin{algorithmic}[1]
        \REQUIRE $L$, $W_{d,n}^{DL}$, $W_{d,n}^{UL}$, $\mathcal{D}_n$, $B_d^n$, $|\Upsilon|$, $f_s$, $f_d^n, \forall I_n \in \mathcal{I}$;
        \ENSURE $\hat{\alpha_n}$, $\mu_{d,n}^{DL}$, $\mu_{d,n}^{UL}$, $\theta_d^n$;
        \STATE $\hat{a}_n\leftarrow 0.5, \mu_{d,n}^{DL}\leftarrow 1/N, \mu_{d,n}^{UL}\leftarrow 1/N, \theta_d^n\leftarrow 1/N, \forall I_n \in \mathcal{I}$;
        \WHILE{\textit{TRUE}}
            \STATE $\hat{\alpha}_n \leftarrow$ solving the problem $\mathcal{P}_2$ with fixed $\mu_{d,n}^{DL}$, $\mu_{d,n}^{UL}$ and $\theta_d^n$;
            \STATE $\mu_{d,n}^{DL} \leftarrow$ solving the problem $\mathcal{P}_2$ with fixed $\hat{\alpha}_n$, $\mu_{d,n}^{UL}$ and $\theta_d^n$;
           \STATE $\mu_{d,n}^{UL} \leftarrow$ solving the problem $\mathcal{P}_2$ with fixed $\hat{\alpha}_n$, $\mu_{d,n}^{DL}$ and $\theta_d^n$;
           \STATE $\theta_d^n \leftarrow$ solving the problem $\mathcal{P}_2$ with fixed $\hat{\alpha}_n$, $\mu_{d,n}^{DL}$ and $\mu_{d,n}^{UL}$;
           \STATE $Q_j \leftarrow \sum_{n=1}^{N}\tau_n^t(l_n,\mu_{d,n}^{DL},\mu_{d,n}^{UL},\theta_d^n)$
           \IF{$|(Q_j-Q_{j-1})/Q_j|<\sigma$}
              \RETURN
           \ENDIF
           $j=j+1;$
        \ENDWHILE
        \STATE $\hat{a}_n=arg\min|a_n-\hat{a}_n|, a_nL\in{1,2,....,L}$
    \end{algorithmic}   
\end{algorithm}

\begin{algorithm}[ht]
    \caption{Decentralized proactive computing resource and spectrum resource allocation scheme}
    \label{The de-ppcs algorithm}
    \begin{algorithmic}[1]
        \REQUIRE $L$, $\mu_{d,n}^{DL}$, $\mu_{d,n}^{UL}$, $\theta_d^n$;
        \ENSURE $\hat{\alpha}_n$
        \STATE Initialize a count $j=1$;
        \STATE Initialize $\mu_{d,n}^{DL}\leftarrow 1/|N|$, $\lambda_n\leftarrow 0$, $z_n\leftarrow 0, \forall I_n\in I$;
        \WHILE{$\lVert\nabla\mu_{d,n}^{DL}(j)\rVert_2$+$\lVert\nabla\lambda(j)\rVert_2$+$\lVert\nabla z(j)\rVert_2$ $>$ $\sigma$}
            \STATE The edge server broadcasts the $\lambda_n(j)$, $z_n(j), \forall I_n\in \mathcal{I}$ to other end devices, $\mathcal{A}_n$;
            \STATE These $N$ end devices update and get $\mu_{d,n}^{DL}(j+1)$, $\lambda_n(j+1)$,$z_n(t+1)$;
           \STATE These $N$ end devices transfer $\lambda_n(j+1)$, $z_n(j+1)$ to the edge server;
           $j=j+1;$
        \ENDWHILE
    \end{algorithmic}   
\end{algorithm}

\section{Theoretical analysis}
\subsection{The convergence and optimality of subproblems}
\textbf{Lemma 1. } \emph{The problem $\mathcal{P}_3$ is strictly convex with respect to $\mu_{d,n}^{DL}$}.

\emph{Proof.} For any feasible $\mu_{d,n}^{DL}$, $\mu_{d,m}^{DL}$, $\forall I_n, I_m \in I$, we have
\begin{equation}
    \frac{\partial^2 Q}{\partial \mu_{d,n}^{DL}\partial \mu_{d,m}^{DL}}=\left\{
    \begin{aligned}
        0, n\neq m,\\
        \sum_{\upsilon \in \Upsilon}\sum_{b\in D_d^n}\frac{2\psi_{s,n}^{g,tr}(l_n)}{({\mu_{d,n}^{DL}})^3R_{DL}}, n=m
    \end{aligned}
    \right.
\end{equation}
where $R_{DL}=W^{DL}log_2(1+P_{s}|h_d^n|^2/W^{DL}N_0)$ , $2\psi_{s,n}^{g,tr}(l_n)$ is positive and ${({\mu_{d,n}^{DL}})^3R_{DL}}>0$. Thus, the Hessian matrix $\mathbf{H}=(\frac{\partial^2 Q}{\partial \mu_{d,n}^{DL}\partial \mu_{d,m}^{DL}})_{N\times N}$ is symmetric and positive definite. Constraint $C_2$ is an affine function with respect to $\mu_{d,m}^{DL}$. Constraints $C_1$, $C_3$ and $C_4$ are irrelevant to $\mu_{d,m}^{DL}$. Therefore, the problem $\mathcal{P}_3$ is strictly convex with respect to $\mu_{d,m}^{DL}$.

\textbf{Lemma 2. } \emph{The problem $\mathcal{P}_3$ is strictly convex with respect to $\mu_{d,n}^{UL}$}.

\emph{Proof.} For any feasible $\mu_{d,n}^{UL}$, $\mu_{d,m}^{UL}$, $\forall I_n, I_m \in I$, we have
\begin{equation}
    \frac{\partial^2 Q}{\partial \mu_{d,n}^{UL}\partial \mu_{d,m}^{UL}}=\left\{
    \begin{aligned}
        0, n\neq m,\\
        \sum_{\upsilon \in \Upsilon}\sum_{b\in D_d^n }\frac{2 \psi_{d,n}^{s,tr}(l_n)}{({\mu_{d,n}^{UL}})^3 R_{UL}} , n=m
    \end{aligned}
    \right.
\end{equation}
where $R_{UL}=W^{UL}log_2(1+P_d^n|h_d^n|^2/W^{UL}N_0)$ , $2(\psi_{d,n}^{s,tr}(l_n))$ is positive and ${({\mu_{d,n}^{UL}})^3 R_{UL}}>0$. Thus, the Hessian matrix $\mathbf{H}=(\frac{\partial^2 Q}{\partial \mu_{d,n}^{UL}\partial \mu_{d,m}^{UL}})_{N\times N}$ is symmetric and positive definite. Constraint $C_2$ is an affine function with respect to $\mu_{d,m}^{DL}$. Constraints $C_1$, $C_3$ and $C_4$ are irrelevant to $\mu_{d,m}^{UL}$. Therefore, the problem $\mathcal{P}_3$ is strictly convex with respect to $\mu_{d,m}^{UL}$.

\textbf{Lemma 3. } \emph{The problem $\mathcal{P}_3$ is strictly convex with respect to $\theta_d^n$}.

\emph{Proof.} For any feasible $\theta_d^n$, $\theta_d^m$, $\forall I_n, I_m \in I$, we have
\begin{equation}
    \frac{\partial^2 Q}{\partial \theta_d^n\partial \theta_d^m}=\left\{
    \begin{aligned}
        0, n\neq m,\\
        \sum_{\upsilon \in \Upsilon}\sum_{b\in D_d^n }\frac{2\phi_{s,n}^{b,e}(l_n)}{(\theta_d^n)^3 f_s}, n=m
    \end{aligned}
    \right.
\end{equation}
where ${2B_d^n\phi_{s,n}^{b,e}(l_n)}$ is positive, and ${(\theta_d^n)^3 f_s}>0$. Thus, the Hessian matrix $\mathbf{H}=(\frac{\partial^2 Q}{\partial f_s^n\partial f_s^m})_{N\times N}$ is symmetric and positive definite. Constraint $C_4$ is an affine function with respect to $f_s^n$. Constraints $C_1$, $C_3$ and $C_4$ are irrelevant to $f_s^n$. Therefore, the problem $\mathcal{P}_3$ is strictly convex with respect to $f_s^n$.

\textbf{Proposition 1. } \emph{According to \cite{yi2016initialization}, Algorithm 2 can convergence to the global optimal solution with any initial condition if the following conditions are satisfied:}
\begin{itemize}
    \item The function $Q(\mu_{d,n}^{DL}),\forall I_n\in I$, are continuously differentiable convex functions;
    \item There exist feasible solutions for the optimization problem $\mathcal{P}_2$;
    \item The information exchange among end devices can be formulated as an undirected and connected graph.
\end{itemize}

\emph{Proof.} Proposition 1 gives three conditions for Algorithm 2 to converge to the global optimal solution with any initial condition. If we can prove that these conditions hold for Algorithm 2, then we prove that Algorithm 2 can converge to the global optimal solution with any initial condition. In the following, we prove that these three conditions hold.

\begin{itemize}
    \item According to the lemma 1, since the Hessian matrix of the function $Q(\mu_{d,n}^{DL}),\forall I_n\in I$ with respect to $\mu_{d,n}^{DL}$ and $\mu_{d,m}^{DL}$ is symmetric and positive definite, the function $Q(\mu_{d,n}^{DL}),\forall I_n\in I$ are continuously differentiable convex functions. Hence, the first condition in proposition 1 holds for Algorithm 1.
    \item {According to \emph{Lemma 1-3}, it can be observed that problem $\mathcal{P}_3$ is strictly convex with respect to $\mu_{d,n}^{DL}$ , $\mu_{d,n}^{UL}$, and $\theta_d^n$. Therefore, problem $\mathcal{P}_2$ is also strictly convex and has a feasible solution. Hence, the second condition in proposition 1 holds for Algorithm 1.}
    \item Since $N$ end devices are all connected to the edge server and the edge server can relay the information among end devices, the information exchange among these end devices can be modeled as an undirected and connected graph with assistance of the edge server. Hence, the third condition in proposition 1 holds for Algorithm 1.
\end{itemize}

So far, we have proved that these three conditions in proposition 1 hold. Thus, we have proved that Algorithm 2 can converge to the global optimal solution with any initial condition.

\subsection{The convergence of global problem}
\textbf{Proposition 2. } \emph{Algorithm 2 based on the BCD method is proposed to solve the global optimization problem including four subproblems. According to \cite{grippo2000convergence}, Algorithm 2 based on the BCD method can converge if at least two of the subproblems are strictly quasi-convex}. 

\emph{Proof.} Suppose that there exists $x_1$, $x_2$ and $\lambda$, where $\lambda \in (0,1)$, such that the strictly convex function is not strictly quasi convex function, we can derive

\begin{equation}
    f(\lambda x_1+(1-\lambda)x_2)\geq \mathop{max}\{f(x_1),f(x_2)\}
\end{equation}

It is obvious that the inequalities $\mathop{max}\{f(x_1),f(x_2)\}\geq f(x_1)$ and $\mathop{max}\{f(x_1),f(x_2)\}\geq f(x_2)$ always hold. As a result, we can derive

\begin{equation}\label{contradictsStrictlyConvex}
\begin{aligned}
   f(\lambda x_1+(1-\lambda)x_2)\geq \lambda\mathop{max}\{f(x_1),f(x_2)\}+ \\
   (1-\lambda)\mathop{max}\{f(x_1),f(x_2)\}\geq \lambda f(x_1)+(1-\lambda)f(x_2)
\end{aligned} 
\end{equation}

We can find that Eq. \ref{contradictsStrictlyConvex} contradicts the definition of strictly convex, where $f(\lambda x_1+(1-\lambda)x_2)<\lambda f(x_1)+(1-\lambda)f(x_2)$, such that our above assumption is not true. As a result, there are three strictly quasi-convex subproblems. Hence, the condition for Algorithm 2 converging is satisfied. We thus complete the proof. Moreover, the subproblem respect with $\alpha_n$ is not convex. Therefore, the global problem does not necessarily converge to a global optimal solution, but a local optimal solution.

In section 6, we have proved that the optimization problem $\mathcal{P}_1$ is strongly convex with respect to $\mu_{d,n}^{DL}$, $\mu_{d,n}^{UL}$ and $f_s^n$. We prove strictly convex function to be strictly quasi convex function by adopting the contradiction.

\section{Performance evaluation}
In this section, we conduct extensive experiments to evaluate the performance of the proposed DP-MORA scheme. We first introduce the related experimental settings including IoT-edge computing environment parameters, training datasets, DNN models and several baseline algorithms. Then, we verify the efficiency and accuracy of the proposed DP-MORA scheme. Finally, we compare the DP-MORA scheme with these baseline approaches under different experimental settings, and analyze their impacts on the DP-MORA scheme.

\subsection{Experimental settings}

\emph{\textbf{IoT-Edge Computing Environment:}} The IoT-edge computing environment is consisting of an edge server and 10 end devices. By default, the computing capacity of the edge server is set to 60GFLOPS. The radio spectrum bandwidths for the edge server's downlink and uplink are set to 50Mbps and 100Mbps, respectively. As end devices' resource capacities are heterogeneous, we set three types of end devices: (1) Raspberry Pi3 with 3.62GFLOPS CPU-cycle frequency and 4GB memory in total, denoted as rpi3; (2) Raspberry Pi-3A+ with 5.0GFLOPS CPU-cycle frequency and 4GB memory in total, denoted as rpi3A+; (3) Raspberry Pi-4B (4GB) with 9.69GFLOPS CPU-cycle frequency and 4GB memory in total, denoted as rpi4B. These 10 end devices consist of 4 rpi3s, rpi3A+s and rpi4Bs. These 10 end devices can communicate with the edge server over the wireless network. 

\emph{\textbf{Training DataSet:}} Our experiments are conducted on two real-world image classification datasets: (1) \emph{CIFAR-10 dataset} \cite{krizhevsky2009learning}. There are 60,000 32$\times$32 colourful images, of which 50000 images are used for model training samples and 10000 images are used for model evaluation samples. These 60,000 colourful images are classified into 10 classes, with 6000 images per class. Each colourful image is labeled as one of ten classes, such as "Cat" or "Dog". (2) \emph{MNIST dataset }\cite{lecun1998gradient}. It contains around 70,000 28$\times$28 grayscale images of handwritten digits which 60000 images are used for model training samples and 10000 images are used for model evaluation samples. Each grayscale image is labeled as one of ten classes of handwritten digits from "0" to "9".


\emph{\textbf{DNN Models:}} In an IoT-edge computing environment, the edge server cooperates with these heterogeneous end devices to train a global DNN model. Referring to the literature \cite{geiping2020inverting}, we select ResNet 18 \cite{he2016deep} and ResNet 34 two models as the evaluation models. That is because the sample data can be recovered partially based on the gradients of server-side model \cite{geiping2020inverting}. The ResNet 18 consists of a convolution (CONV) layer, a max-pooling (POOL) layer, eight BasicBlocks and a fully-connected (FC) layers. The first two, the second two, the third two and the fourth two of these eight BasicBlocks are two $3*3*64$, two $3*3*128$, two $3*3*256$, and two $3*3*512$ convolution operations, respectively. The ResNet 34 consists of a convolution (CONV) layer, a max-pooling (POOL) layer, sixteen BasicBlocks and a fully-connected (FC) layers. The first three, the next four, the next six and the next three of these sixteen BasicBlocks are three $3*3*64$, four $3*3*128$, six $3*3*256$, and three $3*3*512$ convolution operations, respectively.

\emph{\textbf{Benchmarks:}} We first select four typical split Federated Learning schemes:

(1) FedAvg \cite{mcmahan2017communication}: The full model on each end device is trained in parallel and their model parameters are aggregated to the edge server to obtain the updated global model.

(2) SplitFed1 \cite{thapa2022splitfed}: The full model on each end device is splited at the same cut layer. The identical server-side sub-model is offloaded to the edge server from each end device for device-edge synergy training in a sequential way.

(3) SplitFed2: It adopts the same model offloading strategy as ours, and multiple end devices train different sub-models in a sequential manner.

(4) FederSplit \cite{turina2020combining}: Each end device offloads the identical server-side sub-model to the edge server for device-edge synergy training in a parallel manner.

(5)SplitFed3: It adopts the same model offloading strategy as ours, but multiple end devices train different sub-models in a parallel manner.

Existing researches mainly focus on identifying the optimal cut layers for end devices. They fail to consider the impact of the
edge server’s computation and network resource allocation
on the training efficiency. In this paper, we jointly consider the problem of cut layer selection and resource allocation. Since there is no related resource allocation policy, we select two typical resource allocation schemes:

(1) Average Fair (AF): It stands for the average resource allocation, which equally allocates the computing resource of the edge server and the communication resource to $N$ end devices.

(2) Proportional Fair (PF): It is the abbreviation of proportional fair, which allocates the computing resource of the edge server and the communication resource according to the mini-batch sizes of $N$ end devices.

Finally, we combine them to obtain eight joint offloading and resource allocation strategies: FedAvg+AF(FAAF), SplitFed1+PF (SF1PF), SplitFed1+AF(SF1AF), SplitFed2+PF(SF2PF), SplitFed2+AF (SF2AF), FederSplit+PF(FSPF), FederSplit+AF(FSAF), SplitFed3+PF(SF3PF), SplitFed3+AF(SF3AF). To demonstrate the proposed DP-MORA scheme's superiority in training efficiency while satisfying the data leakage risk rate constraint, we compare the DP-MORA scheme with these above baseline algorithms.

\subsection{Performance evaluation of the proposed DP-MORA scheme}

\begin{figure}[h]
    \centering
    \includegraphics[scale=0.31,trim=50 0 4 0]{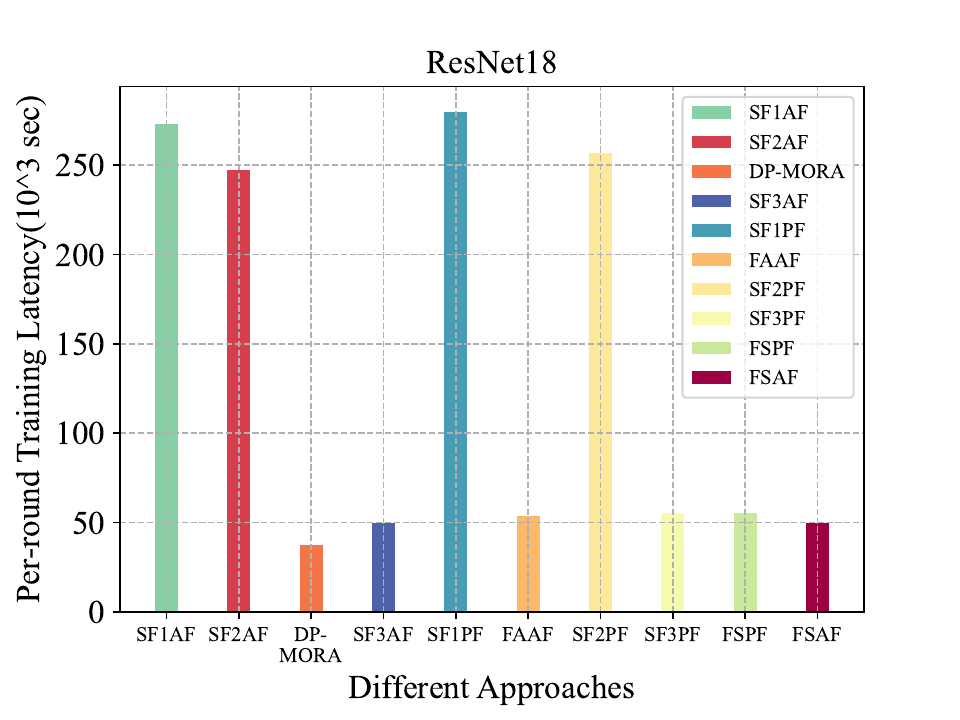}
    \includegraphics[scale=0.31,trim=50 0 50 0]{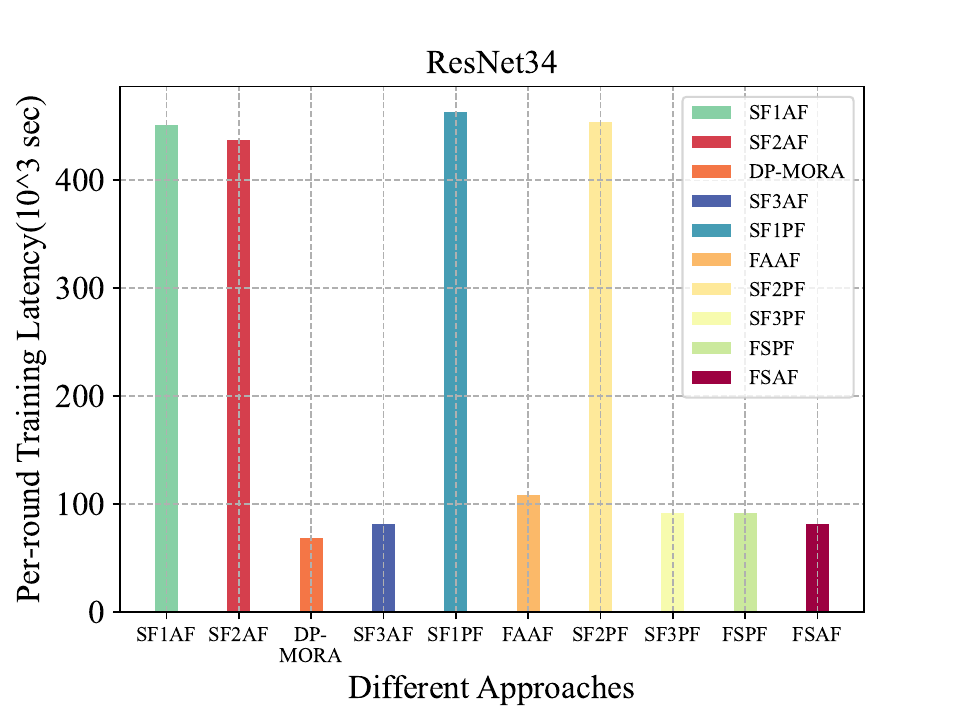}
    
    \caption{Per-round training latencies of different approaches over ResNet18 (left) and ResNet34 (right) when data leakage risk rate constraint is 0.5}
    \label{fig:per_round_training_latency_0_5}
\end{figure}

\emph{\textbf{1) Training latency:}} Fig. \ref{fig:per_round_training_latency_0_5} shows the per-round training latency of different schemes over ResNet 18 and ResNet 34 two models when the data leakage risk rate constraint $P^{risk}$ is 0.5. We can see from Fig. \ref{fig:per_round_training_latency_0_5} that the per-round training latency of the DP-MORA scheme is $24.95\%$ lower than that of SF3AF, $24.09\%$ lower than that of FAAF, $31.72\%$ lower than that of SF3PF and FSPF, $86.02\%$ lower than that of SF1AF, $86.35\%$ lower than that SF1PF, $84.56\%$ lower than that of SF2AF, $85.14\%$ lower than that SF2PF, and $24.09\%$ lower than that of FSAF when the data leakage risk rate is less than or equal to 0.5. The per-round training latency consists of device-side model distribution latency, training latency of five epochs and device-side model transmission latency. That is because considering the heterogeneity of end devices’ resource capacities and their local data samples’ size, the DP-MORA scheme solves the optimal model offloading and resource allocation scheme which makes the standard deviation of per-epoch training latencies of end devices as small as possible. Therefore, the DP-MORA scheme can mitigate the straggler effect and reduce the per-epoch training latency, thereby reducing the per-round training latency.

The per-round training latency of SF1AF and SF1PF is the highest. There are three main reasons: (1) SF1AF and SF1PF split the DNN model on each end device at the same layer without considering the heterogeneity of end devices' resource capacities and their local data samples’ size; (2) SF1AF and SF1PF collaborate multiple end devices with the edge server to train DNN model in a sequential way, which incur high training latency; (3) SF1AF and SF1PF allocate the shared network resource and computing resource of the edge server in average and proportional way without considering the impact of efficient resource allocation on the training latency. 

The per-round training latency of SF2AF and SF2PF is the second highest. There are two main reasons: (1) although they adopt the same model offloading strategy as the DP-MORA scheme, SF2AF and SF2PF collaborate multiple end devices with the edge server to train DNN model in a sequential way; (2) SF2AF and SF2PF allocate the shared network resource and computing resource of the edge server in an average and proportional way without considering the impact of efficient resource allocation on the training latency.

The per-round training latency of SF3AF and SF3PF is higher than that of the DP-MORA scheme. The main reason is that although SF3AF and SF3PF adopt the same model offloading strategy as the DP-MORA scheme, they do not consider the heterogeneity of end devices' resource capacities and their local data samples' size, but also optimal allocation of bandwidth and computing resources.

FSAF and FSPF offload as many layers as possible to the edge server while satisfying the data leakage risk rate constraint 0.5. However, the per-round training latency of FSAF and FSPF is higher that of the DP-MORA scheme. That is because FSAF and FSPF split each end device's DNN model at the same cut layer without considering the heterogeneity of end devices' resource capacities and their local data samples’ size. In addition, FSAF and FSPF also do not consider the impact of the edge server's resource allocation on the training latency.

FAAF locally trains the whole DNN model on resource-constrained end devices. Therefore, its data leakage risk rate is the lowest. However, the per-round training latency of FAAF is higher than that of the DP-MORA scheme. That is because FAAF does not fully exploit the resource-adequate
edge server to achieve device-edge synergy training, thereby incurring higher per-round training latency.   

\emph{\textbf{2) Waiting latency:}} After multiple epochs of training of each end device, its latest device-side model is transmitted to the edge server to further perform model aggregation. As a result, the earliest end device with its device-side model transmitted to the edge server needs to wait for the latest end device. For the sake of convenience, the waiting latency of an end device is defined as the duration between its device-side model uploaded to the edge server and the last end device's model uploaded to the edge server. Table. \ref{Waiting_latency_resnet18} and Table. \ref{Waiting_latency_resnet34} illustrate each end device's waiting latency of different schemes over ResNet 18 and ResNet 34 two models when the data leakage risk rate constraint $P^{risk}$ is 0.5. Based on Table. \ref{Waiting_latency_resnet18}, the per-round waiting latency variances of SF1AF, DP-MORA, SF2AF, SF3AF, FSAF and FAAF six schemes over ResNet 18 are 6427.60, 78.18 5211.61, 171.48, 123.32, 211.14, respectively. Based on Table. \ref{Waiting_latency_resnet34}, the per-round waiting latency variances of SF1AF, DP-MORA, SF2AF, SF3AF, FSAF and FAAF six schemes over ResNet 18 are 17568.76, 192.118189, 16475.28, 380.02, 326.48, 856.96, respectively. Therefore, the end devices' waiting latencies in DP-MORA fluctuate slightly. That is because the DP-MORA scheme can identify the optimal model offloading and resource allocation strategy. This strategy allocates less bandwidth resources and computing resources of edge server to powerful end devices with fewer samples. Conversely, it allocates more bandwidth resources and computing resources of edge server to those weak end devices with more samples, thereby reducing the waiting latency for less capable end devices with more local data samples and improving the training efficiency. Otherwise, if the bandwidth resource and computing resource of the edge server are equally divided to each end device, the weaker end device with more local data samples needs to spend more time in training models, inevitably incurring more waiting latency and the total training latency, thereby reducing the training efficiency.

We can also see from Table. \ref{Waiting_latency_resnet18} and Table. \ref{Waiting_latency_resnet34} that the waiting latency of the DP-MORA scheme is lowest, the waiting latency of SF3AF, FSAF and FAAF are medium, the waiting latency of SF1AF and SF2AF is highest. The reason for the DP-MORA scheme with lowest waiting latency is that the DP-MORA scheme can identify the optimal model offloading and resource allocation scheme considering the heterogeneity of end devices’ resource capacities and their local data samples’ size. The reason for FSAF and FAAF with the highest waiting latency is that they collaborate multiple end devices with the edge server to train DNN model in a sequential way, thereby incurring high training latency. The reason for SF3AF, FSAF and FAAF with the medium waiting latency is that they train DNN model in a parallel manner. Specifically, SF3AF adopts the same model offloading strategy as the DP-MORA scheme, but it does not consider the impact of efficient resource allocation on training latency. FSAF offloads the identical server-side sub-model to the edge server for device-edge synergy training in a parallel manner. However, it not only fails to consider the heterogeneity of end devices’ resource capacities and their local data samples' size but also fails to consider the edge server's resource allocation. FAAF trains the whole DNN model on local end devices in parallel. However, it does not fully exploit the resource-adequate edge server to achieve efficient device-edge synergy training.

\begin{table}[h]
  \centering

\caption{Per-round waiting latency of each end device over \\ ResNet 18 model}
    \normalsize 
  \resizebox{\linewidth}{!}{ %

    \begin{tabular}{c|cccccc}
    
    \toprule
    
    \multirow{2}{*}{End Devices} & \multicolumn{6}{c}{Approaches}\\

    \cmidrule(l){2-7}
    & SF1AF & \textbf{DP-MORA} & SF2AF & SF3AF & FSAF & FAAF \\
    
    \midrule
    $I_0$ & 0 & \textbf{20.05} & 0 & 29.38 & 29.38 & 31.67 \\
$I_1$ & 20.99 & \textbf{15.96} & 20.99 & 29.1 & 26.66 & 32.46 \\
$I_2$ & 44.69 & \textbf{26.8} & 42.26 & 39.94 & 33.92 & 44.16 \\
$I_3$ & 61.15 & \textbf{5.67} & 52.69 & 8.39 & 8.39 & 9.05 \\
$I_4$ & 103.12 & \textbf{11.73} & 94.66 & 24.84 & 21.92 & 28.09 \\
$I_5$ & 131.57 & \textbf{27.96} & 120.19 & 41.1 & 35.74 & 45.28 \\
$I_6$ & 146.19 & \textbf{0.0} & 129.46 & 0.0 & 0.0 & 0.0 \\
$I_7$ & 196.56 & \textbf{15.96} & 179.82 & 29.1 & 26.66 & 32.46 \\
$I_8$ & 220.27 & \textbf{25.64} & 201.09 & 38.78 & 32.09 & 43.03 \\
$I_9$ & 238.55 & \textbf{10.36} & 212.68 & 15.39 & 15.39 & 16.59 \\
    
    \bottomrule
    
    \end{tabular}
  }
  \label{Waiting_latency_resnet18}
\end{table}

\begin{table}[h]
  \centering

\caption{Per-round waiting latency of each end device over \\ ResNet 34 model}
  \resizebox{\linewidth}{!}{ %
    \normalsize 
    \begin{tabular}{c|cccccc}
    \toprule
    
    \multirow{2}{*}{End Devices} & \multicolumn{6}{c}{Approaches}\\

    \cmidrule(l){2-7}
    & SF1AF & \textbf{DP-MORA} & SF2AF & SF3AF & FSAF & FAAF \\
    
    \midrule
    $I_0$ & 0 & \textbf{34.58} & 0 & 48.28 & 48.28 & 63.81 \\
$I_1$ & 34.49 & \textbf{32.71} & 34.49 & 43.53 & 43.53 & 65.39 \\
$I_2$ & 73.72 & \textbf{41.98} & 73.72 & 59.85 & 55.17 & 88.96 \\
$I_3$ & 101.32 & \textbf{9.52} & 96.64 & 13.79 & 13.79 & 18.23 \\
$I_4$ & 170.29 & \textbf{27.38} & 165.61 & 35.69 & 35.69 & 56.59 \\
$I_5$ & 217.37 & \textbf{44.56} & 212.69 & 62.4 & 58.23 & 91.23 \\
$I_6$ & 241.91 & \textbf{0.0} & 233.06 & 0.0 & 0.0 & 0.0 \\
$I_7$ & 324.68 & \textbf{32.71} & 315.82 & 43.53 & 43.53 & 65.39 \\
$I_8$ & 363.91 & \textbf{39.43} & 355.06 & 57.31 & 52.1 & 86.69 \\
$I_9$ & 394.58 & \textbf{17.64} & 380.52 & 25.29 & 25.29 & 33.43 \\
    
    \bottomrule
    
    \end{tabular}
  }
  \label{Waiting_latency_resnet34}
\end{table}

\begin{figure}[h]
    \centering
    \includegraphics[scale=0.12,trim=45 0 85 20,clip]{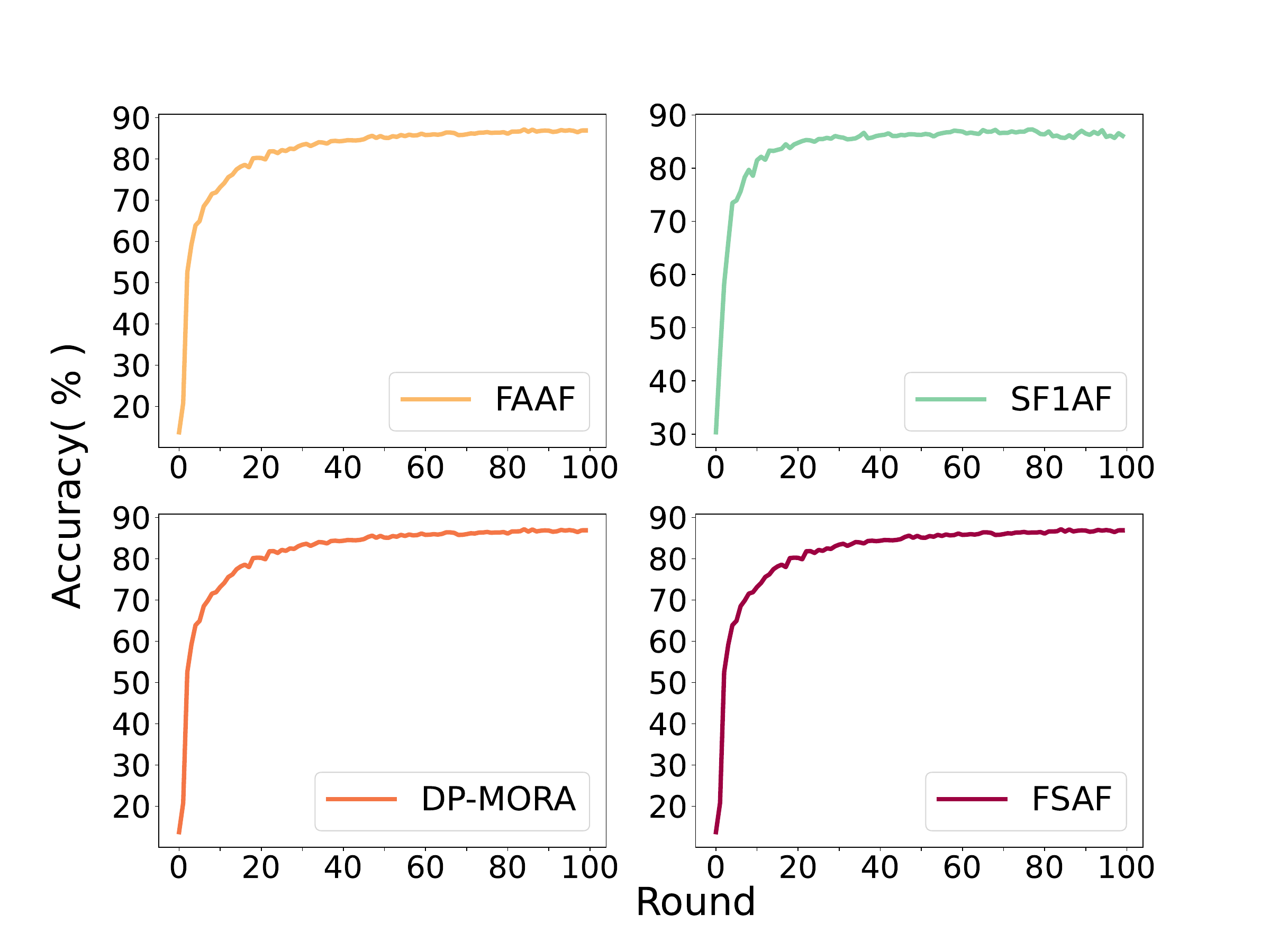}
    \includegraphics[scale=0.3,trim=10 0 40 30,clip]{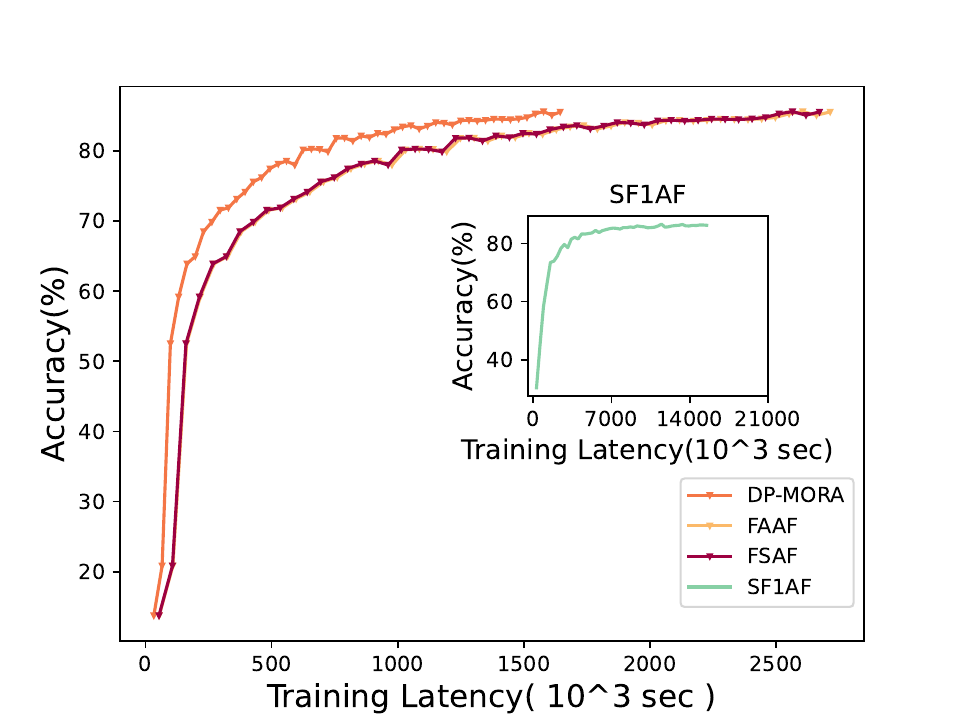}
    \caption{Accuracies of different approaches over ResNet 18 on cifar10 dataset}

    \label{fig:risk_rate_resnet18}
\end{figure}

\begin{figure}[h]
    \centering
    \includegraphics[scale=0.12,trim=45 0 85 20,clip]{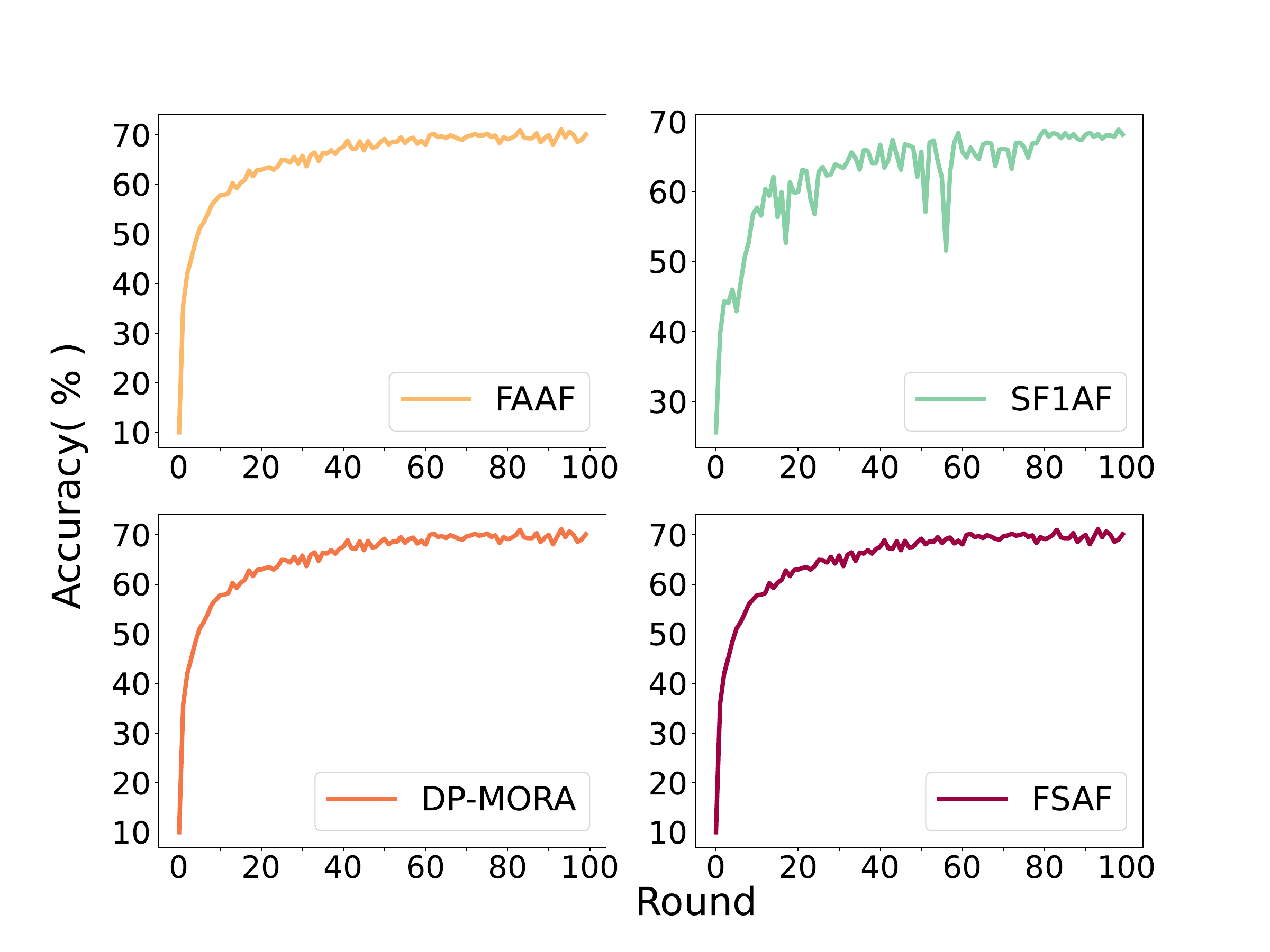}
    \includegraphics[scale=0.3,trim=10 0 40 30,clip]{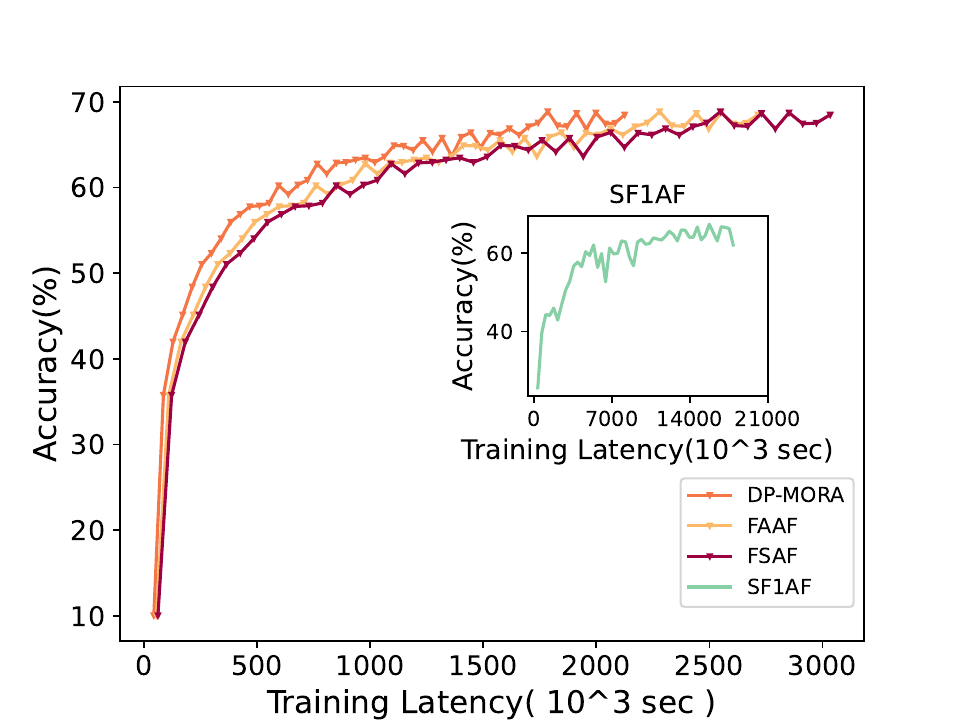}
    \caption{Accuracies of different approaches over ResNet 34 on cifar10 dataset}

    \label{fig:risk_rate_resnet34}
\end{figure}

\emph{\textbf{3) Model accuracy:}} To verify the impact of the proposed DP-MORA scheme on the model accuracy, we plot its model accuracy curves over ResNet 18 and ResNet 34 with the increase of the training rounds and the training latency. Fig. \ref{fig:risk_rate_resnet18} and Fig. \ref{fig:risk_rate_resnet34} illustrate the model accuracy of different schemes over ResNet 18 and ResNet 34 two models, when the data leakage risk rate constraint $P^{risk}$ is 0.5. We can observe from Fig. \ref{fig:risk_rate_resnet18} and Fig. \ref{fig:risk_rate_resnet34} that the model accuracy of the DP-MORA scheme gradually increases and eventually converges with the increase of the training rounds. In addition, we can observe that the model accuracy of the DP-MORA scheme is consistent with that of FAAF. The main reason is that the DP-MORA scheme mainly optimizes on training efficiency by jointly model offloading and resource allocation, which has no influence on model accuracy.

Moreover, we can observe that the DP-MORA scheme takes a shorter training latency than FAAF, FSAF and SF1AF benchmarks to reach convergence. Specifically, the time consumed by the DP-MORA scheme to reach convergence is about 1250x$10^3$  seconds, while those consumed by FAAF, and FSAF are about 2000 x $10^3$ seconds. SF1AF takes 12000x$10^3$ seconds to reach convergence. That is because the per-round training latency of the DP-MORA scheme is lower than those of FAAF, SF1AF and FSAF. The overall training latency is the product of the per-round training latency and the number of training rounds. Therefore, the proposed DP-MORA scheme can converge faster than all benchmarks.

\begin{figure}[h]
    \centering
    \includegraphics[scale=0.30,trim=20 0 40 0,clip]{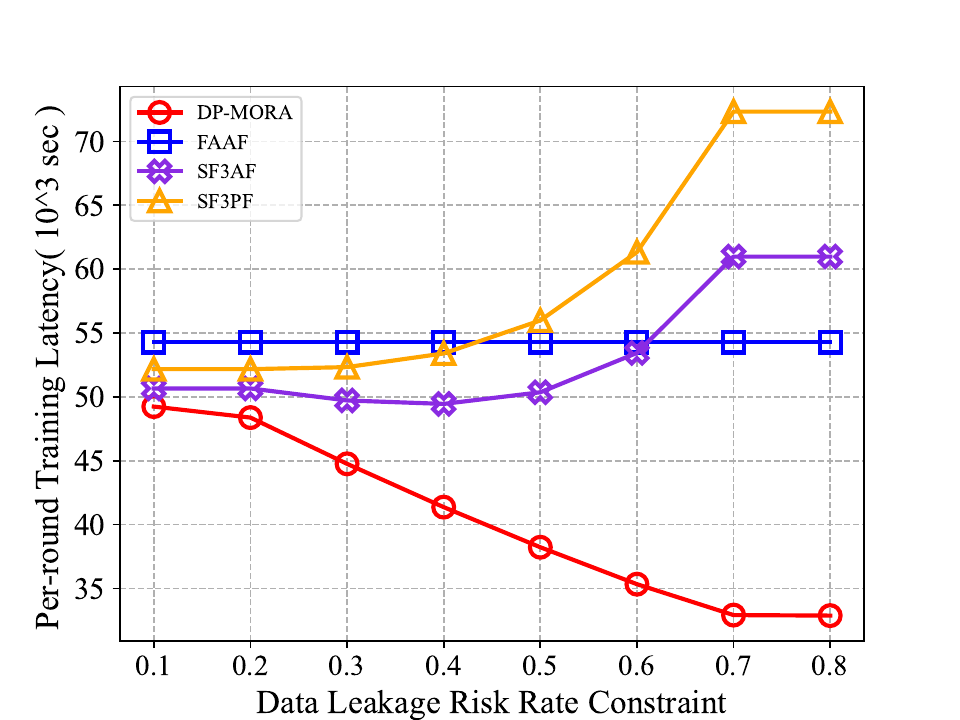}
    \includegraphics[scale=0.30,trim=15 0 45 0,clip]{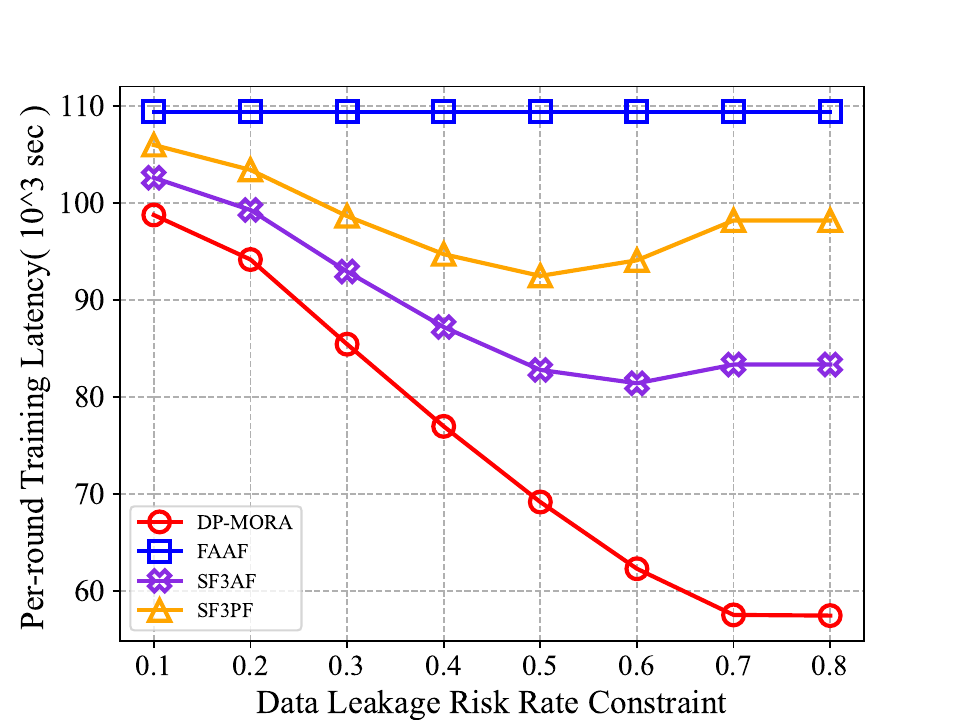}
    \caption{Per-round training latency of different approaches over ResNet 18 (left) and ResNet 34 (right) with different data leakage
risk rate constraints}

    \label{fig:risk_rate}
\end{figure}

\emph{\textbf{4) Impact of different data leakage risk rate constraints:}} To examine the impact of different data leakage risk rate constraints on the total training latency, we vary the data leakage risk rate constraint from 0.1 to 0.8 with the increment of 0.1. Fig. \ref{fig:risk_rate} plots the related experimental results. In Fig. \ref{fig:risk_rate}, we can see that the total training latency of the DP-MORA scheme gradually decreases as the data leakage risk rate constraint increases. That is because different data leakage risk rate constraints correspond to different cut layers requirement. The higher the data leakage risk rate constraint is, the shallower the cut layer is, thereby the solution space of the optimization problem with minimizing training latency while satisfying the data leakage risk rate constraint is larger. The DP-MORA scheme has a higher probability of identifying the optimal solution from the larger feasible solution space. The optimal solution is the optimal cut layer and resource requirements for each end device. The optimal solution splits the DNN model at an optimal shallow cut layer and offloads the server-side sub-model with more layers to the resource-adequate edge server, and allocates optimal bandwidth resource for the transmission data and optimal computation resource for the server-side sub-model, thereby obtaining lower training latency. On the contrary, when the data leakage risk rate constraint is low, it indicates that end devices have a low tolerance for data leakage. Based on this, only several deep-cut layers of the DNN model can be selected to meet the data leakage risk rate constraint. A deep cut layer means heavy computation workloads on resource-constrained end devices and light computational workload on resource-adequate edge server, which incur a higher training latency.

Moreover, we can further observe that compared to SF3AF, SF3PF and FAAF benchmarks, the DP-MORA scheme achieves the lowest total training latency while satisfying the data leakage risk rate constraints. Specifically, when the data leakage risk rate constraint is 0.8, compared to SF3AF, SF3PF and FAAF, the DP-MORA scheme reduces the total training latency of the ResNet18 model up to $24.95\%$, $31.72\%$ and $24.09\%$, respectively. That is because the DP-MORA scheme identifies the optimal model offloading and resource allocation strategy for heterogeneous end devices to reduce the total training latency.

\begin{figure}[h]
    \centering
    \includegraphics[scale=0.28,trim=0 0 30 0,clip]{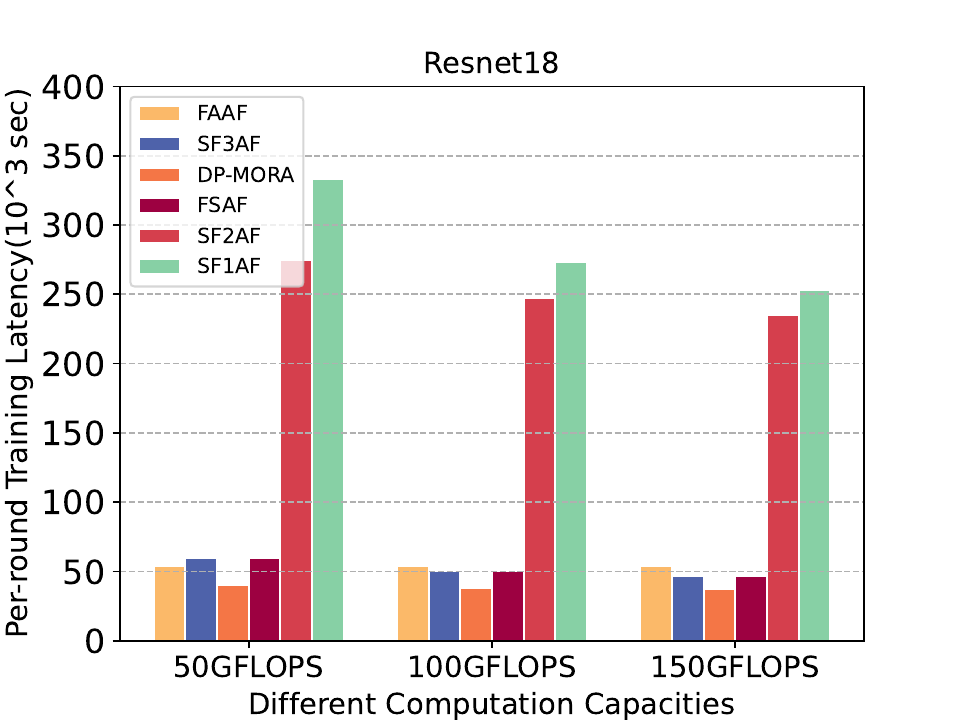}
    \includegraphics[scale=0.28,trim=0 0 30 0,clip]{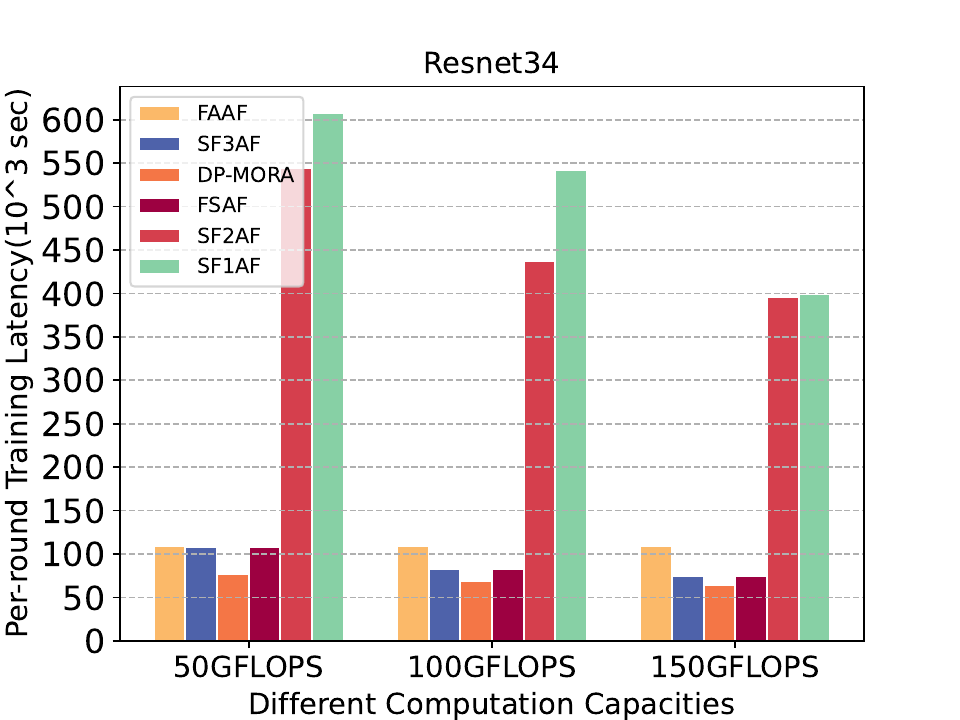}
    \caption{Per-round training latency of different computation capacities}
    \label{fig:risk_rate_com}
\end{figure}

\emph{\textbf{5) Impact of edge server's different computation capacities:}} We investigate the impact of edge server's different computation capacities on the training latency. We vary the computing capacity of the edge server from 50GFLOPS to 150GFLOPS with the increment of 50. The related experimental result is given in Fig. \ref{fig:risk_rate_com}. We can see from Fig. \ref{fig:risk_rate_com} that when the data leakage risk rate constraint $P^{risk}$ is 0.5, the per-round training latency of DP-MORA, SF3AF, FSAF, SF2AF and SF1AF schemes over ResNet 18 and ResNet 34 two models gradually decrease with the increase of the edge server's computation capacity. The main reason is that the edge server with the higher computation capacity can cooperate end devices to handle more workloads, and thereby reducing the per-round training latency. The per-round training latency of FAAF over ResNet 18 and ResNet 34 two models are constant with the increase of the edge server's computation capacity. That is because FAAF trains the whole DNN model on local end devices in parallel. Therefore, the increase of the edge server's computation capacity has no effect on the training latency of FAAF.  

We can further observe from Fig. \ref{fig:risk_rate_com} that the training latency of the DP-MORA scheme is lower than those of other baseline algorithms. Specifically, when the computation capacity is 150GFLOPS, the per-round training latency of the DP-MORA scheme is $31.5\%$ lower than that of FAAF, $21.1\%$ lower than that of FSAF and SF3AF, $84.1\%$ and $85.3\%$ lower than that of SF2AF and SF1AF when the data leakage risk rate constraint 0.5 is satisfied on ResNet18 model. And when the computation capacity is 50GFLOPS, the per-round training latency of the DP-MORA scheme is $26\%$ lower than that of FAAF, $33.1\%$ lower than that of SF3AF and FSAF, $85.4\%$ lower than that of SF2AF, $87.9\%$ lower than that of SF1AF when the data leakage risk rate constraint 0.5 is satisfied on ResNet18 model. That is because the edge server's computation resource becomes more and more adequate with the increase of the edge server's computation capacity. The DP-MORA scheme enables to take full use of more adequate computation resources to collaborate with multiple heterogeneous end devices to efficiently train DNN model, thereby greatly reducing the training latency.

\begin{figure}[h]
    \centering
    \includegraphics[scale=0.28,trim=0 0 30 0,clip]{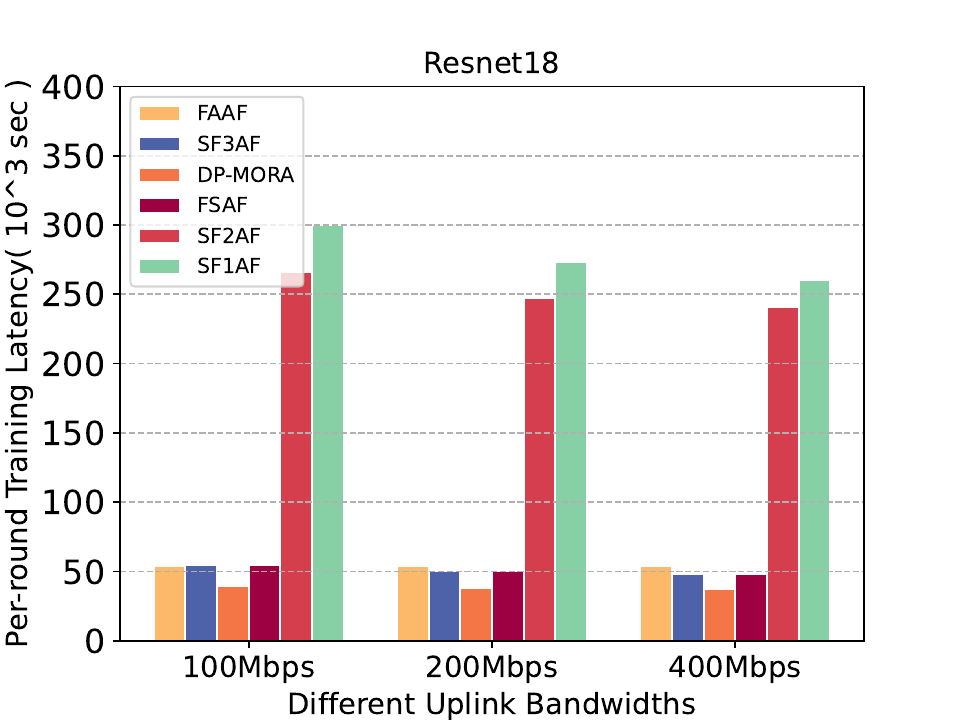}
    \includegraphics[scale=0.28,trim=0 0 30 0,clip]{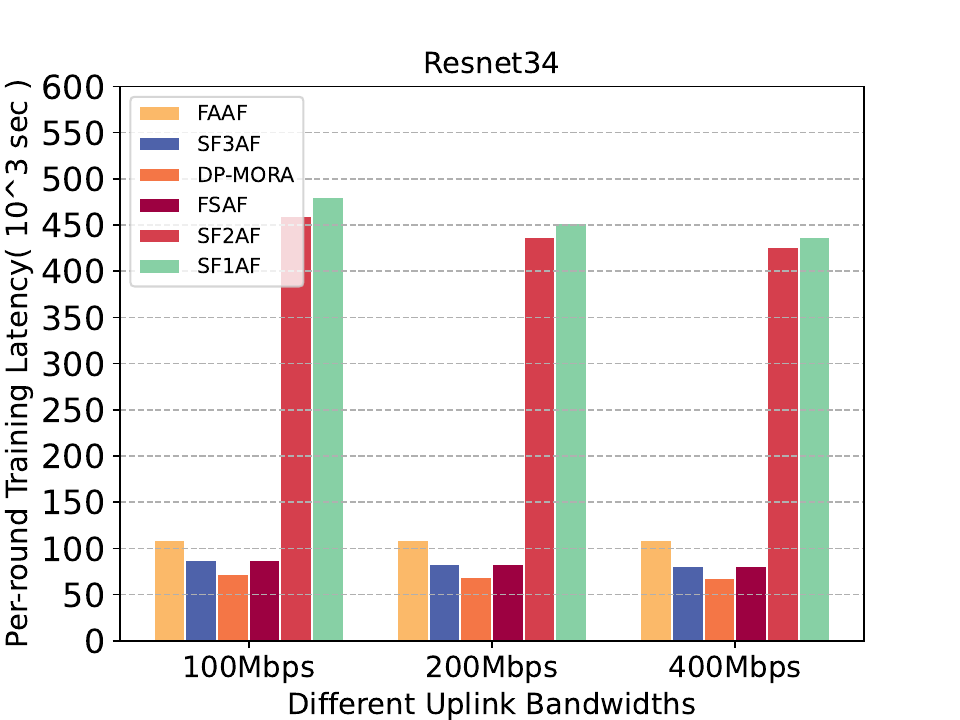}
    \caption{The per-round training latency of different uplink bandwidths}
    \label{fig:risk_rate_up}
\end{figure}

\emph{\textbf{6) Impact of edge server's uplink bandwidth:}}
We evaluate the impact of edge server's uplink bandwidth on the training latency. We vary the edge server's uplink bandwidth from 100Mbps to 400Mbps. The related experimental result is shown in Fig. \ref{fig:risk_rate_up}. We see that the per-round training latency of the DP-MORA scheme and other baseline algorithms gradually decrease. That is because when the edge server's uplink bandwidth increases, the time taken to transmit the same amount of data can decrease. Moreover, when the edge server's uplink bandwidth increases, we can also see that the per-round training latency of the proposed DP-MORA scheme is always lower than that of other benchmarks. Specifically, when the edge server's uplink bandwidth is 100Mbps, the DP-MORA scheme reduces the per-round training latency by $36.6\%$, $27.4\%$, and $27.4\%$ as compared with the FAAF, SF3AF, and FSAF benchmarks, respectively. The DP-MORA scheme is $85\%$ lower than that of SF2AF, and $86.7\%$ lower than that of SF1AF due to the sequential execution instinct for SF1AF and SF2AF. When the edge server's uplink bandwidth is 400Mbps, the DP-MORA scheme reduces the per-round training latency by $31.2\%$, $22.5\%$, and $22.5\%$ as compared with the FAAF, SF3AF, and FSAF benchmarks, respectively.
DP-MORA is $84.5\%$ lower than that of SF2AF, and $85.6\%$ lower than that of SF1AF. That is because the DP-MORA scheme optimally allocates edge server's uplink bandwidth resources for multiple heterogeneous end devices, thereby reducing the training latency.

\begin{figure}[h]
    \centering
    \includegraphics[scale=0.28,trim=0 0 30 0,clip]{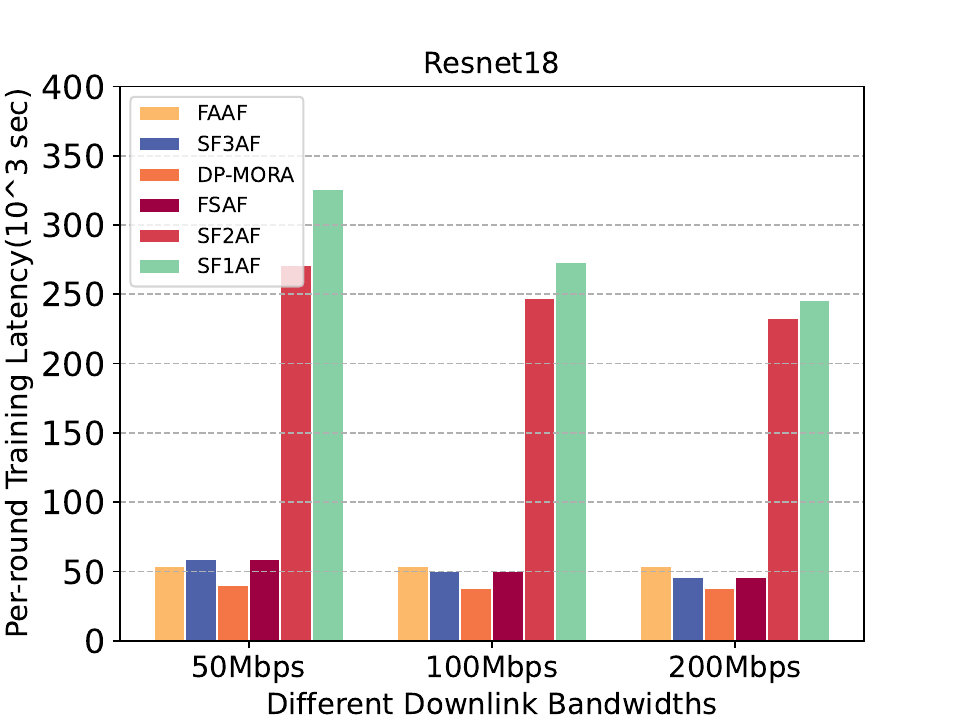}
    \includegraphics[scale=0.28,trim=0 0 30 0,clip]{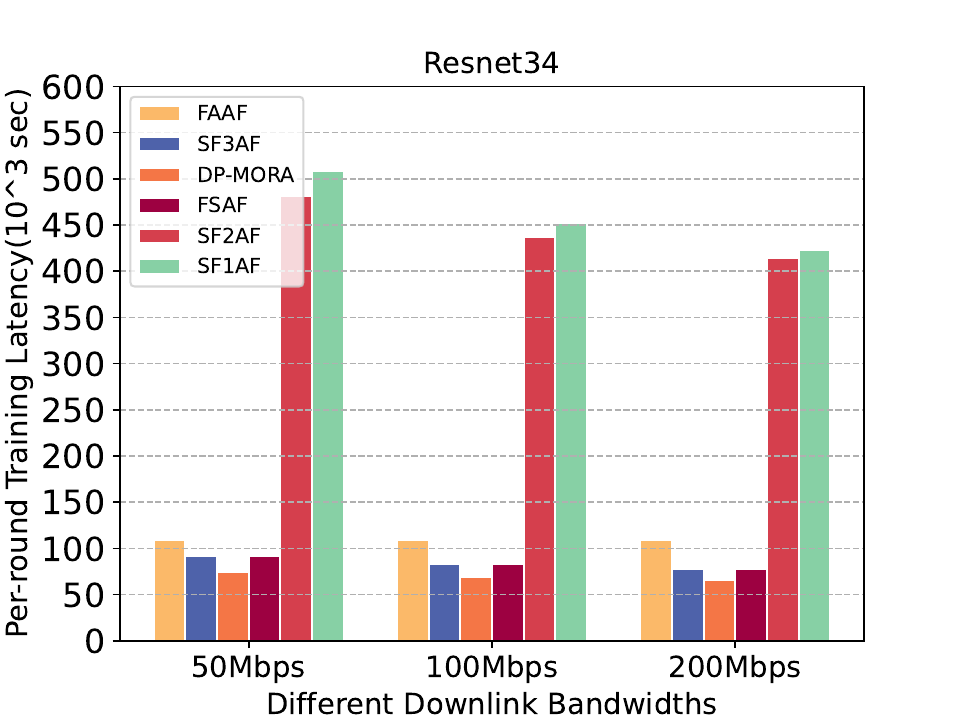}
    \caption{The per-round training latency of different downlink bandwidths}
    \label{fig:risk_rate_down}
\end{figure}

\emph{\textbf{7) Impact of edge server's downlink bandwidth:}} 
In order to examine the impact of edge server's downlink bandwidth on the per-round training latency, we vary the edge server's downlink bandwidth from 50Mbps to 200Mbps.  The related experimental results are shown in Fig. \ref{fig:risk_rate_down}. We see that when the edge server's downlink bandwidth increases, the per-round training latency of the DP-MORA scheme and other benchmarks decreases. The main reason is that it takes less time to transmit the same amount of data with the larger uplink bandwidth. Moreover, we can further see from Fig. \ref{fig:risk_rate_down} that the per-round training latency of the DP-MORA scheme is lower than that of other benchmarks. Specifically, when the edge server's downlink bandwidth is 50Mbps on the ResNet18 model, the DP-MORA is $25.6\%$ lower than that of FAAF, $31.4\%$ lower than that of SF3AF and FSAF, $85.1\%$ lower than that of SF2AF, $87.6\%$ lower than that of SF1AF. When the edge server's downlink bandwidth is 200Mbps, the DP-MORA is $29.9\%$ lower than that of FAAF, $17.5\%$ lower than that of SF3AF and FSAF, $83.6\%$ lower than that of SF2AF, $84.5\%$ lower than that of SF1AF. That is because considering the heterogeneity of end devices’ resource capacities and their local data samples’ size, the DP-MORA scheme can optimally allocate the downlink bandwidth resource to each end device.

\section{Conclusion}
In this paper, we investigate the training efficiency of split federated (SplitFed) learning while satisfying the data leakage risk rate constraint in resource-constrained IoT-edge computing environment. To address this problem, we first formulate the latency of DNN model training and measure the data leakage risk rate of DNN model training adopting split federated learning. Then we formulate joint model offloading and resource allocation problem to be a mixed integers non-linear programming. At last, we design a decentralized and privacy-preserving joint model offloading and resource allocation scheme to optimize the per-round training latency while satisfying data leakage risk rate constraint. We conduct extensive experiments on two real-world datasets to evaluate the performance. Extensive experiments show that the DP-MORA scheme can effectively reduce the per-round training latency while satisfying the data leakage risk rate constraint.

\section*{Acknowledgments}
This work was supported by the National Natural Science Foundation of China (Nos. 62202133, 62125206), the Zhejiang Provincial Natural Science Foundation of China (No. LY23F020015), the Key Research Project of Zhejiang Province under Grant 2022C01145.

\bibliographystyle{IEEEtran}
\bibliography{refeerences.bib}


\begin{IEEEbiography}[{\includegraphics[width=1in,height=1.25in,clip,keepaspectratio]{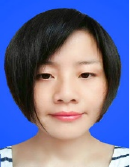}}]{Binbin Huang}
is an Assistant Professor in the College of Computer Science at the University of Hangzhou Dianzi, in Hangzhou, China. She received her PhD degree in Computer Science and Technology from Beijing University of Posts and Telecommunications in 2014. His research interests include distributed artificial intelligence, cloud computing, mobile edge computing, reinforcement learning and workflow scheduling.
\end{IEEEbiography}
\vspace{-1cm}

\begin{IEEEbiography}[{\includegraphics[width=1in,height=1.25in,clip,keepaspectratio]{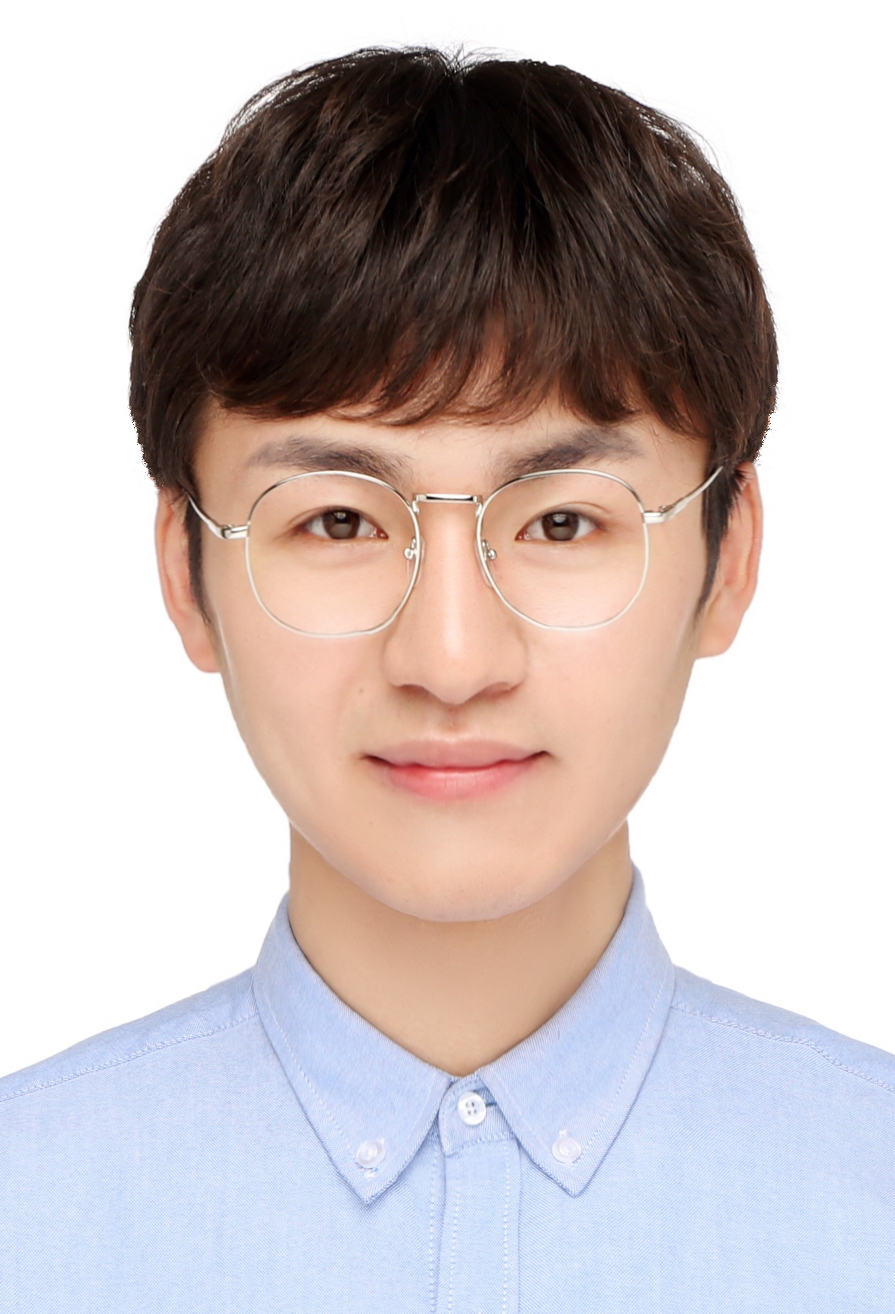}}]{Hailiang Zhao} 
received the B.S. degree in 2019 from the school of computer science and technology, Wuhan University of Technology, Wuhan, China. He is currently pursuing the Ph.D. degree with the College of Computer Science and Technology, Zhejiang University, Hangzhou, China. His research interests include cloud \& edge computing, distributed computing and optimization algorithms. He has published several papers in flagship conferences and journals including IEEE ICWS 2019, IEEE TPDS, IEEE TMC, etc. He has been a recipient of the Best Student Paper Award of IEEE ICWS 2019. He is a reviewer for IEEE TSC and Internet of Things Journal.
\end{IEEEbiography}
\vspace{-1cm}

\begin{IEEEbiography}[{\includegraphics[width=1in,height=1.25in,clip,keepaspectratio]{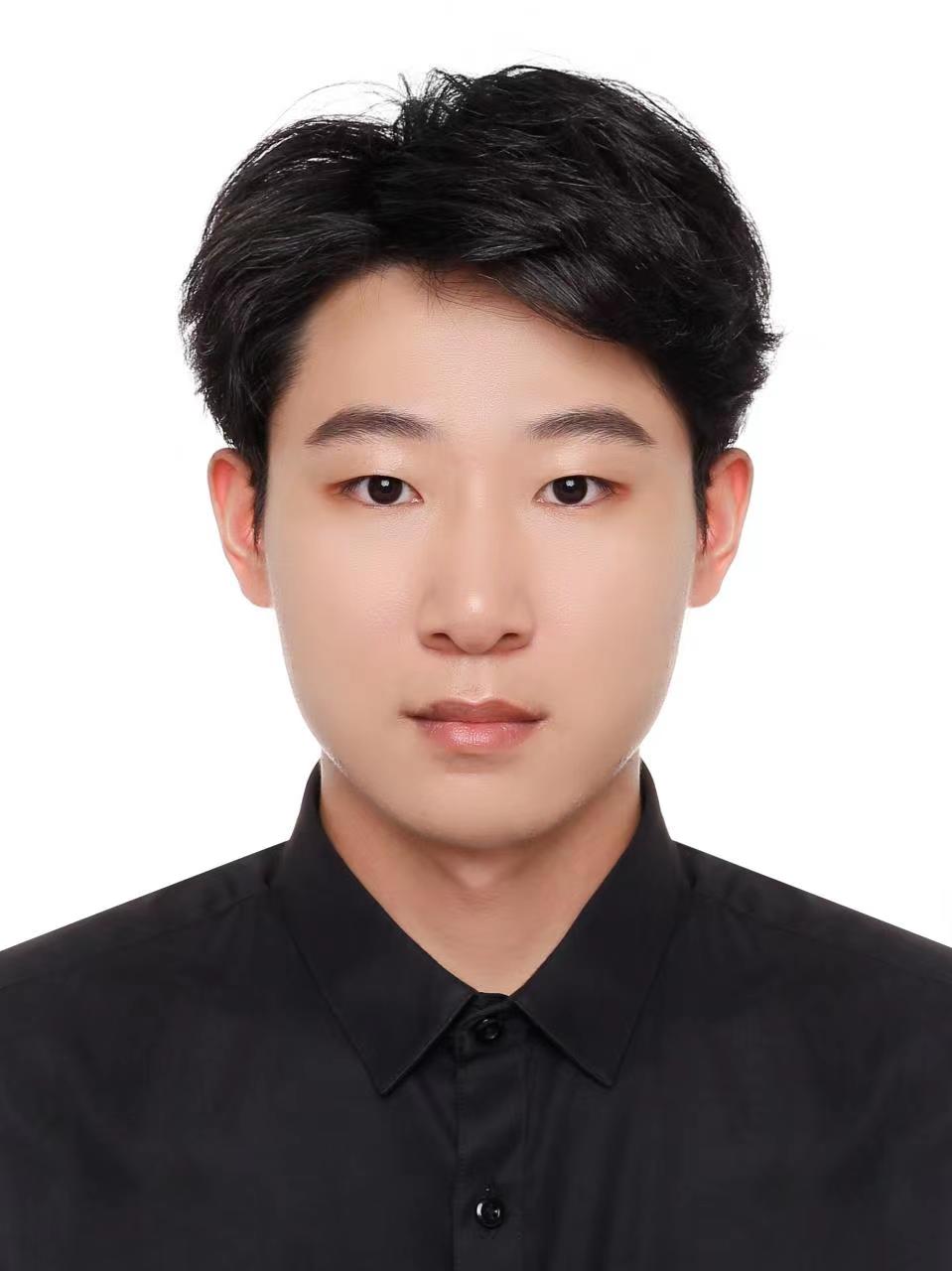}}]{Lingbin Wang} 
received the B.S. degree from Zhejiang Gongshang University, Zhejiang, China 2022. Currently, he is pursuing his M.S degrees in Hangzhou Dianzi University, Zhejiang, China. His research interests include deep learning,
game theory and mobile edge computing.
\end{IEEEbiography}
\vspace{-1cm}

\begin{IEEEbiography}[{\includegraphics[width=1in,height=1.25in,clip,keepaspectratio]{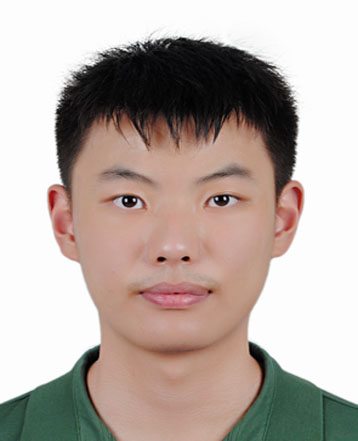}}]{Wenzhuo Qian} 
received the B.S. degree from the School of Computer Science and Technology, Hangzhou Dianzi University, Hangzhou, China, in 2023. He is currently working toward the master's degree in the Polytechnic Institute, Zhejiang University, Hangzhou, China. His research interests include edge computing and service computing.
\end{IEEEbiography}
\vspace{-1cm}

\begin{IEEEbiography}[{\includegraphics[width=1in,height=1.25in,clip,keepaspectratio]{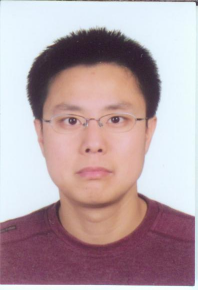}}]{Yuyu Yin}
Yuyu Yin received the PhD in Computer Science from Zhejiang University in 2010. He is currently an Associate Professor at the College of Computer in Hangzhou Dianzi University. His research interests include Service Computing, Cloud Computing, and Business Process Management. He is a IEEE Member, Senior Member of the China Computer Federation (CCF), CCF Service Computing Technical Committee Member. He worked as guest editor for Journal of Information Science and Engineering and International Journal of Software Engineering and Knowledge Engineering, and as reviewers for IEEE transaction on Industry Informatics, Journal of Database Management, Future Generation Computer Systems, ect. 
\end{IEEEbiography}
\vspace{-1 cm}

\begin{IEEEbiography}[{\includegraphics[width=1in,height=1.25in,clip,keepaspectratio]{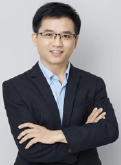}}]{Shuiguang Deng}
is a full professor at the College of Computer Science and Technology in Zhejiang University. He received the BS and PhD both in Computer Science from Zhejiang University in 2002 and 2007, respectively. His research interests include Service Computing, Mobile Computing, and Edge Computing. Up to now he has published more than 100 papers in journals such as IEEE TOC, TPDS, TSC, TCYB and TNNLS, and refereed conferences. He is the Associate Editor of the journal IEEE Trans. on Services Computing and IET Cyber-Physical Systems\: Theory \& Applications. He is a senior member of IEEE.
\end{IEEEbiography}
\vspace{-1 cm}

\vfill

\end{document}